# *AN ACCURATE, EASY TO USE ABUNDANCE SCALE FOR GLOBULAR CLUSTERS BASED ON 2.2 μm SPECTRA OF GIANT STARS*

*Jay A. Frogel[1], Andrew W. Stephens[1],*
*Solange Ramírez[1,2], and Darren L. DePoy[1]*

Department of Astronomy, The Ohio State University
140 West 18th Avenue, Columbus, Ohio  43210
e-mail:  {frogel,stephens}@astronomy.ohio-state.edu

---

# ABSTRACT

We present a new method for the determination of [Fe/H] for globular clusters. This new method is based on moderate resolution (~1500) near-IR spectroscopy in the K-band of 6 to 10 of the brightest giants in a cluster. Our calibration is derived from spectra of 105 stars in 15 globular clusters. From measurements of the equivalent widths of three features in these spectra, EW(Na), EW(Ca), and EW(CO), we are able to reproduce the Zinn & West (1984) abundance scale as updated by Harris (1996) to better than ±0.10 dex for clusters with near solar [Fe/H] down to an [Fe/H] of −1.8. A quadratic fit to the data is somewhat better than a simple linear fit. Three advantages of this method are that it can be used for metal rich, heavily reddened globulars in crowded fields, it does not require any knowledge of any other cluster or stellar parameters such as reddening, distance, or luminosity, and it requires only minimal telescope time.

If stellar $(J–K)_0$ and $M_K$ values are available as well, the accuracy of the [Fe/H] estimate is further improved. If observations - either spectra alone or spectra plus colors and magnitudes - of only three stars in a cluster are available, the resulting value of [Fe/H] will be nearly as reliable as that based on two to three times as many stars.

The accuracy of an [Fe/H] value based on observations of CO absorption alone is significantly less than that which results from the three spectroscopic indices for two main reasons. First, the CO bands are saturated for bright giants in high metallicity clusters. Second within a cluster there is considerable intrinsic star-to-star scatter in EW(CO) due to differences in mixing histories on the giant branch. In contrast, there is no evidence for any intrinsic scatter in the Na or Ca indices. Nevertheless, we predict that observations from space of CO absorption in the integrated light of stellar systems will prove to be of great value for abundance determinations at distances as far as the Coma cluster of galaxies.

Finally, a comparison of globular cluster abundances based on high resolution spectroscopy with the Zinn & West/Harris [Fe/H] scale - and by inference ours - leads us to conclude that the two are closely linearly related over the entire range of globular cluster abundances, although there appears to be a small, constant offset between the two.

                stars: abundances − stars: Population II
                Galaxy: bulge
                infrared: stars
                techniques: spectroscopic

1. INTRODUCTION

The distribution over metallicity of the Galactic globular cluster system is clearly bimodal with mean values for [Fe/H] of $\approx -0.5$ and $-1.6$ (Freeman & Norris 1981; Zinn 1985; Armandroff & Zinn 1988); the minimum in the frequency distribution is at [Fe/H] $\approx -1.0$. Globular clusters with [Fe/H] $\geq -1.0$ are preferentially located close to the Galactic Center. Three-quarters of the ~32 clusters with [Fe/H] $\geq -0.8$ (based on the 1999 June version of Harris' compilation; see Harris 1996) lie within 16° of the Galactic Center. Not surprisingly, these clusters are heavily reddened with an average *E(B−V)* of 1.0. Nearly half of the total sample of the 32 clusters with [Fe/H] $\geq -0.8$ lies within 10° of the Center with an average *E(B−V)* of 1.3. In addition to being heavily reddened, most of these clusters also lie in crowded fields.

The heavy reddening of most metal rich globular clusters combined with the crowded fields in which they are found has led to them being poorly studied optically. A notable exception has been the work of Ortolani and his collaborators (*e.g.* Ortolani *et al.* 1997 and references therein). They have derived *VI* color-magnitude diagrams (CMDs) for many of the metal rich, heavily reddened bulge globulars and estimated distances and [Fe/H] values from the CMDs. Spectroscopic abundances, though, are sorely lacking for these globulars. For example, of the 23 clusters that lie within 16° of the Galactic Center, only 7 have metallicities - [Fe/H][3] - from observations of the red Ca II triplet (see the compilation of Rutledge Hesser, & Stetson (1997b, hereafter RHS97). For the whole sample of 32 metal rich clusters, only 11 are included in RHS97.

The situation is even worse when one considers the availability of abundances based on high resolution spectra. In their recent derivation of a new metallicity scale for globular clusters based only on clusters with fine abundance analysis, Carretta and Gratton (1997, hereafter CG) have only three clusters with [Fe/H] $> -1.0$ on the Zinn & West (1984, hereafter ZW) scale: 47 Tuc (NGC104), M71 (NGC6838) and NGC 6352. None of these three clusters, though, has [Fe/H] $> -0.70$ according to Harris (1996).

Spectroscopic observations in the near-IR (the *JHK* bands) can circumvent the problems that beset optical observations of the heavily reddened, metal rich Galactic bulge clusters[4]. There are two key reasons for this: extinction in the *K* band is only one-tenth that in *V* and the contrast between the brighter cluster giants and foreground field stars is enhanced, often by as much as 3 to 5 magnitudes. This enhancement arises because in the *V, R, I* bands heavy blanketing by TiO, VO, and other molecules strongly suppresses the continuum. Kuchinski *et al.* (1995), Frogel *et al.* (1995), and Kuchinski & Frogel (1995, hereafter GC1, GC2, and GC3, respectively), have laid some of the groundwork for the use of near-IR *photometric* observations of globular cluster giants and horizontal branch stars in the determination of cluster distances, reddenings, and [Fe/H] (see also Minniti *et al.* (1995) and recent papers by Davidge (2000a,b) and references therein).

---

[3] Unless stated explicitly, [Fe/H] can refer to both a "true" Fe determination or one of the commonly used proxies for Fe. Rutledge, Hesser, & Stetson (1997b) use the Ca II triplet as such a proxy. This point is discussed in the last section of the paper

[4] For example, Origlia *et al.* (1997) have discussed a metallicity scale for globulars based on H band stellar spectra.

The purpose of this paper is to present a method, based on medium resolution ($\lambda/\Delta\lambda \approx$ 1500) near-IR spectroscopy of individual stars, for the determination of accurate abundances for heavily reddened globular clusters. As will be seen, our new method is parsimonious of telescope time, easily reproducible, and capable of application to stars in nearby galaxies with future space missions - for example NGST. We will show that this method is applicable to clusters with [Fe/H] $\geq -1.7$ . We have also begun to extend our technique to obtain abundances from the *integrated* light of globular clusters, both in our own Galaxy and in M31. We are also working to extend it to more complex stellar systems such as spheroidal galaxies and the bulges of spirals.

We use the 2.29 μm band head of CO, and the Na I and CaI atomic line blends to establish an [Fe/H] scale based on globular clusters with good optical abundance determinations. These are three of the strongest absorption features in K band spectra (see also the discussion in Förster Schreiber 2000). Ramírez *et al.* (1997) and Ali *et al.* (1995) have used similar spectroscopic observations in studies of local field giants and dwarfs, respectively. For the most heavily reddened and crowded clusters we would expect that an [Fe/H] scale based on spectroscopic observations will be of higher accuracy than any based on infrared or optical photometric observations alone. For example, the near-IR photometric method based on a measurement of the slope of a cluster's giant branch (GC1, GC3) has the disadvantage that in especially crowded fields or fields with large differential reddening it can be difficult or impossible to accurately determine the locus of the giant branch over the necessary range in luminosity. We have recently used our new spectroscopic technique to determine [Fe/H] values for about 100 field giants in the inner part of the Galactic bulge (Ramírez *et al.* 2000).

Section two of this paper describes the selection of clusters and gives details of the observations and data reduction insofar as they differ from the procedures followed by Ramírez *et al.* (1997, 2000). In section three we describe the analysis techniques. Section 4 is the heart of the paper in which we consider several approaches to the derivation of an abundance scale based on the K band spectra and discuss potential limitations to its applicability. Section 5 discusses the pros and cons of using the CO band by itself as an abundance indicator. Section 6 is a discussion of some final points and summarizes the main conclusions of the paper. In particular, we examine the assertion of RHS97 and CG that there may be a systematic difference between the Zinn & West (1984) scale and the scale based on high resolution spectroscopy, especially at its metal rich end.

In a second paper (Frogel, Stephens & Ramírez 2001, heareafter Paper II), we will apply the method developed here to derive [Fe/H] values for several heavily reddened, poorly studied Galactic bulge clusters and analyze similar data for giants in ω Centauri, the only Galactic globular with significant star to star variations in [Fe/H].

## 2. SELECTION OF CLUSTERS AND STARS

### 2.1 The Cluster Sample

We have obtained K-band spectra for 129 stars on the upper red giant branches (RGB) of 17 globular clusters. These are clusters with good, optically determined abundances, especially for [Fe/H] $\geq -1.0$; most have previously published *JHK* observations so that we can investigate luminosity and color effects on [Fe/H] determinations. We have also observed bright giants in some lower [Fe/H] clusters to determine the range of applicability of this technique. Table 1 lists the stars we observed. Notes about the clusters in our sample and a brief description of recent, relevant observations of them are contained in the Appendix to this paper.

Ideally, we would have liked to calibrate our NIR spectroscopic indices using clusters with [Fe/H] determined from high dispersion spectra. Unfortunately, as we pointed out in the Introduction, none of the clusters in CG97 have [Fe/H] values greater than −0.6. In fact, Table 3 of RHS97, which summarizes high dispersion spectroscopic analyses of cluster stars, gives values greater than −0.60 for only two clusters, NGC 5927 and 6352. Both of these values were determined by J. G. Cohen (Cohen 1983, and a value quoted in Frogel *et al.*1983a). NGC 6352 *was* studied by CG97, but they derived an [Fe/H] of −0.64, (Cohen's 1983 [Fe/H] for this cluster is −0.37 on a scale which fixes 47 Tuc at −0.70). Therefore, for consistency we decided to use Harris' (1996, with 1999 values on the web) compendium of [Fe/H] values for Galactic globular clusters to calibrate the near-IR indices. Since Harris' compilation is basically a refinement of the ZW values and incorporates more recent data, we will refer to his values as [Fe/H]$_{ZW}$. Indeed, for the 17 clusters for which we have giant star spectra, the mean difference in [Fe/H] between Harris and ZW is −0.02 with a dispersion of ±0.12. The largest difference between the two sets of values is for NGC6528 for which Harris gives −0.17, while the ZW value is +0.12. Table 4 lists the [Fe/H]$_{ZW}$ values from Harris (1996) for the 17 clusters whose giants we have observed spectroscopically.

About half of the clusters we observed have previously published infrared photometric observations (*e.g.* Frogel 1985a; Frogel *et al.* 1979, 1981, 1983b, Kuchinski *et al.* 1995, Kuchinski & Frogel 1985). For most of these clusters we simply chose 6 to 10 of the brightest stars for spectroscopic study. For clusters without available infrared photometric data, as well as for some of those with such data, we took new images of the cluster centers at J and K and chose the brightest stars that appeared to be cluster members based on their location on instrumental color-magnitude diagrams.

## 2.2 Notes on Selection of Stars

For 47 Tucanae we chose two samples of stars from the photometry of Frogel *et al.* (1979). The "bright" sample consists of stars at or close to the top of the RGB. Even though we excluded the 4 known LPVs (V1-4), essentially all of the stars in this sample still vary in brightness to some degree. Thus the correspondence between photometric parameters and the spectral feature strengths will not be as tight as for stars from other clusters. The "faint" sample of 47 Tuc stars was chosen from further down the RGB; on average they are nearly 2 magnitudes fainter in K than stars in the bright sample. As described later, this "faint" sample is used to test the dependence of the equivalent widths of the spectral features on luminosity.

For NGC 288, 362, 4833, and 5927 we chose stars from near the top of the RGB based on the *JHK* photometry of Frogel *et al.* (1983b) and, for NGC 6712, from Frogel (1985a). We observed an additional four previously unobserved stars near the center of NGC 362 that appeared to be comparable in brightness to the stars with photometry. However, these 4 will not be used for the [Fe/H] calibration since their photometry was uncalibrated.

We obtained spectra for 4 infrared bright stars in the inner region of M69, 3 of which are identified in Hartwick & Sandage (1968). Photometry for V1, V3, and V6 is from Frogel & Elias (1988); data for V3 and V6 are also given by Frogel *et al.* (1983b). For the remaining stars, *JK* magnitudes are an average of the values given by Davidge & Simons (1991) and Ferraro *et al.* (1994).

For M71 we chose the brightest giants from Frogel *et al.* (1979) and from GC1. Arp & Hartwick (1971, hereafter AH) 21, 45, and 113 have *JHK* from Frogel *et al.* For AH- I, 64, and 66 the data are from GC1. Their values for AH-H are too bright, probably because of blending with AH 36, based on a comparison of their data with more recent, better resolution images taken at MDM. So for H and 36 we took *K* magnitudes from a snapshot MDM image calibrated with the Frogel *et al.* (1979) and Kuchinski *et al.* (1995) photometry. We used E(*J–K*) = 0.12 and $A_K$ =0.07.

For NGC 6304 we first obtained "snapshot" images with OSIRIS on the CTIO Blanco 4 meter telescope at K. Then we took spectra of the 7 apparently brightest stars in the central part of the cluster. Davidge *et al.* (1992) has observed two of the stars for which we have spectra. Their data for these two stars are included in Table 2. The remaining 5 stars were not observed by them, possibly because they were too bright for their detector.

For NGC 6388 we first took a short exposure *K* band image of the central part of the cluster and then selected 9 stars that were bright at *K* for spectroscopic observation. Afterwards, we noted that three of the 9 stars are known variables (Lloyd Evans & Menzies 1977) with near-IR observations in Frogel & Elias (1988). A true distance modulus of 15.30 was used for this cluster (Harris 1996).

For NGC 6440 we chose the 8 brightest stars from Table 3 of GC3. All of these stars appear to be cluster members based on their location in a *K*, *J-K* color-magnitude diagram.

For NGC 6528 we selected bright stars from uncalibrated *JHK* images of the central part of the cluster. Davidge (2000a and private communication) has *JHK* values for 4 of the stars.

For NGC 6553 we first obtained "snapshot" images in *J* and *K* with the 2.4 m Hiltner telescope at MDM Observatory. We then selected the brightest stars in the central region of the cluster (to minimize the possibility of field star contamination) with *J−K* colors that lay on or close to the RGB ridge line. Davidge & Simons (1994) have published *JH* photometry for this cluster. For V4 and V5 we used photometry from Frogel & Elias (1988). For the other stars we took the values from Guarnieri *et al.* (1998). These values were not converted to the CIT/CTIO system.

For NGC 6624 we chose the 5 brightest stars from Table 2 of GC3. All of these stars appear to be cluster members based on their location in a *K*, *J-K* color-magnitude diagram.

Finally, for M5, M15, and M92 we selected the brightest stars from *JHK* images of the central regions of the clusters. No calibrated photometry is yet available for these stars.

## 3. OBSERVATIONS AND DATA REDUCTION

The observations were obtained over several observing seasons. Most were made on the Blanco 4m telescope at CTIO with both CTIO's near-IR spectrometer (the IRS; R= 1650; DePoy *et al.* 1990) and with the Ohio State InfraRed Spectrograph (OSIRIS; R=1380; DePoy *et al.* 1993). Spectral coverage was generally between 2.17 μm and 2.34 μm. These were the same observing runs with the same instrumental set up during which time we acquired data for our study of red giants in the Galactic bulge (Ramírez *et al.* 2000). The tip-tilt secondary on the Blanco 4m was used on the more recent CTIO runs. Additional data were obtained with TIFKAM on the 2.4 m telescope at MDM Observatory with comparable spectral resolution. Centering the stars in the 0.6" slits of the spectrographs on both telescopes was confirmed by maximizing the signal. Our data reduction procedures and equivalent width measurements are similar to those described in Ramírez *et al.* (1997, 2000).  We summarize the important points here.

Each of our spectra begin as a set of 7-10 individual spectra taken at different positions along the slit to eliminate the effects of uneven slit illumination, bad pixels, and fringing.  These are obtained with a long slit (several arc minutes) which allows for simultaneous star and sky measurements.  A sky frame is constructed for each star by median combining the individual frames in each set, effectively removing all traces of the star itself; this sky frame is then subtracted from each spectral frame.  The result is divided by a dome flat field to eliminate the array response, and each spectra is extracted using the IRAF[5] `apall` routine . These individual spectra are then averaged together to yield the final spectrum.

The wavelength calibration is determined by fitting to approximately 12 atmospheric OH lines (Oliva & Origlia 1992) where the sky spectra are the median combination of a matching number of sky spectra extracted from the sky frame using the program star as an extraction template.

Bright standard stars are observed frequently throughout the night in identical fashion as the program stars.  Generally, we observe a standard as close as possible in time and zenith distance for every object star.  The standards are of middle A type or hotter with no significant spectral features in the observed wavelength regime except for Brackett $\gamma$ at 2.16 $\mu$m . Each extracted and averaged program star spectrum is divided by the appropriate normalized standard star spectrum to correct for telluric absorption features.  The result is then multiplied by a 10,000K blackbody spectrum to restore its over all shape.

The technique we use to measure equivalent widths is fully described by Ramírez *et al.* (1997); note, though, that we have slightly modified some of the wavelength intervals for measurement.  While the Ca and Na EW measurements use the same integration limits as Ramírez *et al.* (1997), the EW(CO) measurement uses a slightly smaller integration region.  For this paper and our recent study of bulge giants (Ramírez *et al.* 2000) the continuum regions are all larger than in the earlier paper eliminating many of the small gaps. These larger regions bias the continuum levels slightly lower (artificially reducing the EW measurements), but by including more pixels, the noise level is reduced, giving more consistent measurements.  Table 2 gives the continuum and line regions over which we integrated.  Note that our line region for the CO absorption feature is somewhat wider than that used by Kleinmann & Hall (1986) or Origlia & Oliva (2000), but still only includes absorption from $^{12}$CO; there is no contribution from the $^{13}$CO band head.

Table 1 lists the stars we observed, the photometric parameters from the sources mentioned above, and the measured equivalent widths of the three spectral features.  Eleven of the stars in Table 1 were observed twice.  Six more giants in other globular clusters (Paper II) were also observed twice.  While in a few cases the two observations were obtained during the same observing run, most repeat measurements were separated by one to six years and were often made with different spectrographs.  Table 3 gives the average differences and the dispersions for the 17 pairs of measurements.  The dispersions are comparable in size to the mean differences and are similar to the values found by Ramírez *et al.*(2000).  We regard these dispersions, then, as good estimators of the total uncertainty in our measurements.

---

[5] IRAF is distributed by the National Optical Astronomy Observatories, which are operated by AURA Inc., under cooperative agreement with the NSF.

## 4. ANALYSIS OF THE LINE INDICES

### 4.1. Photometric Parameters and a Qualitative Look at the Spectroscopic Indices

We have measured near-IR spectral indices (Table 1) for 129 giant stars in 17 globular clusters. For 92 of these stars in 14 of the clusters there are near-IR colors and magnitudes available, mainly from Frogel, Persson & Cohen (1979, 1981, 1983b), Frogel (1985), and Kuchinski & Frogel (1995). Average values of the spectroscopic indices for each cluster along with cluster-average values of the reddening and extinction corrected $J-K$ colors and $M_K$ magnitudes for the stars with spectra are given in Table 4. Values for reddening and distance modulus are from Harris (1996). No photometric data are available for the stars observed spectroscopically in M5, M15, or M92. As examples of our data we illustrate in Figures 1 and 2 the spectra for all stars observed in each of two clusters - NGC 6440 and M71. The short vertical bars indicate the features we have measured. The number in parentheses after the star identification is the star's absolute $K$ magnitude. In order to give a sense of how the spectra change from cluster to cluster, we show in the two parts of Figure 3 unweighted average spectra for all stars in each cluster. Numbers in parentheses are the $[Fe/H]_{ZW}$ values for the clusters.

Figure 4 shows the dependence of EW(CO) on $[Fe/H]_{ZW}$ while Figure 5 shows the dependencies of EW(Na) and EW(Ca) on $[Fe/H]_{ZW}$. These figures illustrate the point that the relationships between the spectroscopic indices and $[Fe/H]_{ZW}$ is not simply linear. The behavior of the Ca index is similar to that of Na but the scatter of EW(Ca) at constant $[Fe/H]_{ZW}$ is larger than that for EW(Na) (there is a larger uncertainty attached to its measurement – see Table 3). For both of these indices the measured EWs are consistent with zero for $[Fe/H] < -1.6$. The maximum sensitivity of EW(CO) to $[Fe/H]_{ZW}$ occurs for clusters in the middle of the metallicity range. Clusters more metal poor than NGC 4833 ($[Fe/H]_{ZW} = -1.8$) have EW(CO) so close to zero that sensitivity to $[Fe/H]$ is lost; the $K$ band spectra of the observed giants in M15 and M92 are essentially featureless (see Fig. 3). This is not to say that there is no CO absorption at these low metallicities, but with the spectral resolution of our observations the lines and bands are undetectable. Furthermore, the sensitivity of EW(CO) to metallicity appears to decrease for clusters with $[Fe/H]_{ZW}$ greater than about $-0.6$. While EW(Na) shows a similar decreasing sensitivity to $[Fe/H]_{ZW}$ at the low end of the scale, over the rest of the observed $[Fe/H]$ range EW(Na) is more nearly linear. The difference in behavior of EW(Na) and EW(CO) is clearly evident in Figure 6.

Figs. 4 - 6 are useful as diagnostic tools. For example, in Figs. 4 and 5 the stars in NGC 6304 appear on average to have CO and Na indices that are too strong for their $[Fe/H]_{ZW}$ while these indices for the stars in NGC 6624 have indices that are a bit weak. However, Fig. 6 shows that the *relative* strengths of the two indices for stars from the two clusters are normal and that NGC6624 appears to be more metal poor than NGC 6304, contrary to previous measurements of the metallicities of these clusters. We also note from Fig. 6 that two stars in NGC6304 have especially strong indices. This could be a sign that they are not cluster members at all but rather belong to the bulge field population. Nevertheless, we will use the $[Fe/H]_{ZW}$ values for both of these clusters in the derivation of the calibration based on all observed stars.

Based on the qualitative trends that the data exhibit in Figs. 4, 5, and 6, we will proceed as follows: We will exclude the two most metal poor clusters, M15 and M92 ($[Fe/H] = -2.2$ and $-2.3$, respectively) from the analysis since none of our measured indices display any sensitivity to metallicity below [Fe/H] ~ -1.8. Over the past two decades the CO absorption in the *K*-band has often been used as a metallicity indicator. With our new data set we can quantitatively evaluate how effective EW(CO) by itself is as a metallicity indicator since it is between 5 and 10 times stronger than either the Na or Ca features (although its relative dispersion appears quite similar) and could thus be valuable in determining [Fe/H] for the distant objects[6].

To minimize the effects of any potential luminosity sensitivity on the equivalent widths, we limit our analysis to stars with $M_K$ brighter than −4.0. We use our observations of two groups of giants in 47 Tucanae to help assess the effects of luminosity on [Fe/H] estimates. The first group is on average 1.8 mags fainter in $M_K$ and 0.23 mags bluer in $(J–K)_0$ than the second group and is purposely not included in the calibration. We also treat the 4 stars from the central region of NGC 362 separately. We are left with 105 stars with spectroscopically measured indices. Of these, 77 also have near-IR colors and magnitudes. Table 4 lists the average values of the 5 photometric and spectroscopic parameters for each cluster based only on the reduced samples of 105 and 77 stars. The actually numbers of stars in each cluster that are used in the analysis are given later in Table 11.

### 4.2. Correlation Between EW(CO) and CO (index)

Fifty-three of the globular cluster giants we observed spectroscopically also have had CO indices measured photometrically by Frogel, *et al.* (1979, 1981, 1983b) and Frogel (1985). This photometric index is a color index measured as the difference between a narrow band continuum filter centered at 2.20 μm and a CO filter at 2.36 μm. Its units are magnitudes. Figure 7 shows that the correlation between the spectroscopic and photometric measures of the CO band strength is excellent. A linear least squares fit to the data gives

$$\text{EW(CO)} = 3.226(\pm 0.69) + 82.1(\pm 5.1) \times \text{CO(index)} . \qquad 1$$

The units of all EW measurements in this paper are Ångstroms. The coefficient of 82 is about twice what one obtains from eqns. 1 and 2 of Origlia & Oliva (2000). This difference arises because we employ a wider band-pass that encompasses more of the CO absorption for our EW measurements. The standard error of a prediction of EW(CO) from the photometric CO index is ±2.31; the R value of this fit is 0.91. There are 4 two-sigma outliers (circled points in Fig. 7) - 47 Tuc - V8, NGC 288 - A260 (=V1), NGC 5927-532, and NGC6388-V1; three of these 4 are large amplitude variables. Without these 4 stars the correlation is

$$\text{EW(CO)} = 3.12(\pm 0.51) + 84.4(\pm 3.8) \times \text{CO(index)} . \qquad 2$$

The R value of this fit is improved to 0.96 while the standard error of a prediction of EW(CO) with this relation is reduced to ±1.68. The uncertainty in EW(CO) is assumed to be ±0.74 (Table 3) for all measurements; the uncertainty in the CO (index) is ±0.018 (Frogel *et al.* 1983b, Table 1). Thus, the standard error of a prediction of EW(CO) from the CO (index) is just that expected from the uncertainties in the two measurements alone, namely 1.70. This means that we can make good estimates of EW(CO) from the existing CO (index) filter data for clusters with no additional spectroscopic observations. The quality of the fit (eqn. 2) implies that even though the filter technique for measuring CO absorption included $^{13}$CO bands, their effect on the measurement is small.

---

[6] Although with ground based observations the CO bands become impossible to observe for redshifts much greater than about z=0.02 due to atmospheric absorption, from space, with NGST for example, no such limitation will apply.

In section 5 we will derive a correlation between EW(CO) and [Fe/H]. Although more limited in usefulness than the relations we will derive that use EW(Na) and NW(Ca) as well, an appropriate selection of filters could yield a large multiplexing advantage in the determination of [Fe/H] for faint objects, especially from space where the effects of thermal background emission and telluric water absorption (just longward of the CO band) are minimized or absent. A potential problem is that EW(CO) appears to saturate for stars from the most metal rich clusters. A way around this problem which we are currently pursuing is to measure spectral features in the integrated light of clusters. Since cool giants contribute a steadily increasing percentage of the total near-IR light as [Fe/H] goes up, the apparent depth of the CO absorption will increase as well in spite of the fact that the strength of the band in the coolest individual stars is saturated. This issue is considered later in the paper.

### 4.3 Correlations Between Spectral Indices and Photometric Parameters

We next examine the correlations between the spectroscopic indices and the photometric colors and magnitudes. As stated earlier, the two metal poor clusters M15 and M92 are excluded from this analysis because of their low metallicities ([Fe/H] < − 2.0). For the 105 stars with near-IR spectral data, the three spectroscopic indices correlate with each other and the [Fe/H]$_{ZW}$ values for their parent clusters at better than the 99.9% level. The *(J–K)$_0$* colors, which have been measured for 77 of the stars, correlate with the indices and the [Fe/H] values at the same high level. While $M_K$ correlates with *(J–K)$_0$* at this level as well, its correlation with [Fe/H] and the spectroscopic indices is not quite as good. The level of correlation of $M_K$ with [Fe/H]$_{ZW}$, EW(Na), EW(Ca), and EW(CO) is 90.2, 99.4, 97.0, and 99.2%, respectively.

We can explain the correlations between the spectral indices and photometric parameters qualitatively. Identical observations of 100 stars in a dozen fields in the Galactic bulge, all of which can be considered to be at the same distance and to have a smaller spread in [Fe/H] than the sample of globular cluster giants, show that these spectral indices have at most a weak dependence on luminosity (Ramírez *et al.* 2000). In an old stellar population with a small age spread the *(J–K)$_0$* color of the RGB shifts strongly to the blue as [Fe/H] decreases since *(J–K)$_0$* tracks temperature and less metal rich RGBs lie at systematically hotter temperatures at constant magnitude (see Fig. 13 in Kuchinski *et al.* 1995 and Frogel *et al.* 1983a). Furthermore, the three spectral indices we use all increase slowly in strength with decreasing T$_{eff}$ (Ramírez et al. 1997). Given these results and the generally small *intra*-cluster dispersions in the reddening corrected values for *J–K* and $M_K$ for the observed stars (see Table 4), the tight correlations between the spectral indices, [Fe/H], and *(J–K)$_0$* is understandable. Furthermore, in a cluster of fixed [Fe/H] the RGB has a steep, well defined slope in a *K, J–K* color-magnitude diagram (Kuchinski *et al.* 1995). Hence, it is not surprising that $M_K$ depends most strongly on *(J–K)$_0$* and only weakly on the spectral indices, especially within a given cluster.

In any one cluster the brightest giants have only a small spread in color (or temperature) as is evident from Table 4. Therefore, for all of the stars in one cluster any observed spread in the spectral indices should primarily be due to a combination of measuring uncertainties, differences in the mixing histories of the giant stars (*e.g.* Kraft 1994), and possibly an intrinsic star-to-star spread in [Fe/H]. For the clusters in the calibrating group we regard the latter possibility as unlikely. We can then compare the intra-cluster dispersions in the spectral indices (Table 4) to the dispersions expected based on measurement uncertainty alone (Table 3). Excluding M15 and M92 because of their low metallicities, the mean intra-cluster EW dispersions for the 15 clusters in the calibrating sample are 0.42, 0.62, and 2.75 for EW(Na), EW(Ca), and EW(CO), respectively. For Na and Ca these mean values are consistent with those expected from the measuring uncertainties alone (0.28 and 0.62 from Table 3). There is no reason to suspect any other cause for the dispersion of the few clusters with the largest values. Furthermore, there is no correlation between the Na and Ca dispersions themselves. For EW(CO) the situation is quite different. The mean intra-cluster dispersion of 2.75 (Table 4) is nearly three times larger than that expected from measurement uncertainties alone (0.74, Table 3). Again, there is no correlation between the intra-cluster dispersions in EW(CO) and those of the other two spectroscopic parameters. We conclude, then, that the stars on the RGB we have observed show a significant and intrinsic scatter in the strength of CO absorption within each cluster. The size of the scatter does not appear to be correlated with the [Fe/H] of the cluster. Nor does there appear to be a significant correlation between the departure of an individual star's EW(CO) from the cluster mean and the departure of its $M_K$ value from the cluster mean. The source of the scatter in EW(CO) is most likely differences in mixing histories from one star to another. Evidence for this interpretation is found in Frogel *et al.'s* (1981) observations of giants in 47 Tucanae. Its giants exhibit a large variation in CO at constant color and a strong anti-correlation between CO and CN strength in the stars.

### 4.4 Principal Component Analysis

With the new spectroscopic observations, we have available 5 parameters for 77 of the observed globular cluster giants and 3 parameters for 105 of the giants. The relative effectiveness of these parameters in characterizing the clusters can be assessed with the aid of Principal Component Analysis (PCA). Deeming (1964) gave one of the earliest descriptions of the application of PCA to problems in astronomy. Faber (1973) and Frogel (1985b) used PCA to analyze multi-parameter sets of observations of galaxies. Recently, Francis & Wills (1999) have presented a succinct introduction to the use of PCA in astronomy.

The mean values (averaged over all stars in all calibrating clusters) and standard deviations of [Fe/H]$_{ZW}$ and the two photometric and three spectroscopic variables under consideration are given in Table 5. In order to carry out a PCA we must first transform the variables for each star by subtracting the means and dividing by the standard deviations. These normalized variables then have zero mean and a standard deviation of one. With 6 variables we solve for 5 eigenvectors. The eigenvalues (the variance due to the associated eigenvector) are given in column 2 of Table 6. A good rule of thumb (*e.g.* Francis & Wills 1999) is that an eigenvector with an eigenvalue greater than unity is most likely significant. Conversely, those much less than unity indicate an eigenvector that is most likely insignificant. The first eigenvalue (Table 6) is 4 times larger than the unit variance of the normalized variables; it also accounts for two-thirds of the total variance. It is therefore highly significant. The second eigenvector has a variance close to unity and accounts for about 16% of the total variance. It is probably significant. The 4 remaining eigenvectors have variances significantly less than unity and comparable to what would be expected from the observational and measurement uncertainties alone.

The components of the 5 eigenvectors are listed in Table 7. V1, with two-thirds of the total variance has the strongest contributions from EW(Na) and [Fe/H]$_{ZW}$ and only a weak component from $M_K$. V2 arises almost entirely from the absolute K magnitude. These results are consistent with the finding from the simple correlation analysis presented above, namely that all of the photometric and spectroscopic parameters are highly correlated with one another except for $M_K$.

To complete this section, Table 8 and Table 9 give the results of a PCA analysis for the spectroscopic parameters alone. These tables show that the first eigenvector is 3.2 times larger than the individual variances and accounts for 81% of the total variance. Again, the largest contribution comes from EW(Na) . The second one, which accounts for 12% of the total variance does not appear to be significant since its size is only half that of any of the individual variances. The third eigenvector is negligibly small.

The main conclusion we draw from the PCA analysis is that the three spectroscopic indices, EW(Na), EW(Ca), and EW(CO), form a one-parameter family that depends almost entirely on [Fe/H] with only a weak dependence on luminosity. We also find that although it would be best to have both photometric and spectroscopic information in deriving the [Fe/H] calibration, spectroscopy alone will suffice albeit with somewhat reduced accuracy. Later, we will quantitatively evaluate the effect of not including magnitudes in the final analysis. Finally, in spite of the fact that EW(Na) is on average only one-seventh as strong as EW(CO) it accounts for more of the variation in [Fe/H]$_{ZW}$ than the latter. The likely origin of this effect is the significant and intrinsic star-to-star scatter of EW(CO) within any one cluster.

### 4.5  Simple Linear Fits to the Data

Our objective in this paper is the derivation of a simple and accurate means of calculating [Fe/H] for a globular cluster from near-IR spectra of its bright giants. Therefore, we first carried out a linear least squares fit of [Fe/H]$_{ZW}$ to the three spectroscopic indices and find (solution LL3, Table 10):

$$[\text{Fe/H}]_{LL3} = 0.182 \times \text{EW(Na)} + 0.0571 \times \text{EW(Ca)} + 0.0273 \times \text{EW(CO)} - 1.663$$
$$(\pm 0.03) \quad\quad (\pm 0.025) \quad\quad (\pm 0.005) \quad\quad (\pm 0.06) \ . \quad\quad 3$$

The standard error of an estimate of [Fe/H] from this equation is ±0.18, while the value of R is 0.91. Note the relative lack of sensitivity of the solution to EW(Ca). Exclusion of EW(Ca) as a variable yields a nearly identical answers. On the other hand, this solution is equally sensitive to EW(Na) and EW(CO)  when allowance is made for the fact that the mean value of the latter index is about 7 times greater than the mean of the former for the 105 stars (Table 1). We then inverted the process and calculated [Fe/H] for each of the 105 stars from the above equation and determined the mean [Fe/H] for each of the 15 calibrating clusters. These values are given in Table 11 (LL3) together with the intra-cluster dispersions of the stellar [Fe/H]$_{LL3}$ values. Figure 8 illustrates the means and dispersions. Columns 3 of  Table 11 indicates how many stars in each cluster were used for the solutions based only on spectroscopic parameters.

It is obviously advantageous to have an [Fe/H] scale that relies only on spectroscopically measured parameters and does not require knowledge of the reddening or distance to the cluster. Determination of which stars are the brightest in a cluster and have a reasonably high probability of cluster membership can be easily done from an uncalibrated color-magnitude diagram.

However, we showed with the PCA that *(J–K)₀* is also a significant contributor to the first eigenvector, and that the second eigenvector, although a factor of ~4 less important, depends strongly on $M_K$, a quantity which correlates only weakly with the spectroscopic features. Therefore, we repeated the simple linear multiple regression analysis and included the two photometric parameters for 77 stars in 13 clusters for which they are known. Column 4 of Table 11 indicates how many such stars are in each cluster.

The regression of the spectral features, colors and absolute magnitudes to [Fe/H]$_{ZW}$ is better than the fit with the spectral features alone. The equation for the fit (LLA9 in Table 10) is

$$[Fe/H]_{LLA9} = 0.202 \times EW(Na) - 0.0251 \times EW(Ca) + 0.0259 \times EW(CO) + 0.749 \times (J-K)_0 + 0.151 \times M_K - 1.451 \quad 4$$
$$(\pm 0.04) \qquad (\pm 0.03) \qquad (\pm 0.005) \qquad (\pm 0.23) \qquad (\pm 0.033) \quad (\pm 0.18)$$

The uncertainty in an estimate for [Fe/H] based on this equation is now ±0.14 while R=0.95. As for the previous solution (LL3), Table 11 presents the results of an inversion of the process - the mean [Fe/H] value and dispersion for each cluster based on the LLA9 solution. Some of this improvement over the previous solution is fortuitous and results simply from a different sample of stars. For example, if we repeat the multiple regression fit to the spectroscopic indices only, using the same set of 77 stars with measured colors and magnitudes, we find that the uncertainty in an estimate of [Fe/H] is ±0.16 with an R value of 0.93. These are to be compared to the values of ±0.18 and 0.91, respectively, for the full set of 105 stars derived above. Thus, inclusion of the photometric parameters results in a modest improvement.

### 4.6 Quadratic Solutions

In the previous section we found that simple linear fits to the spectral line indices and the photometric parameters provide good estimates for [Fe/H]$_{ZW}$ on the scale of Harris (1996). Nonetheless, Figures 4, 5, and 6 show that the dependencies *between* the spectroscopic indices and [Fe/H] is not linear. Therefore, we now explore the efficacy of a quadratic fit to the data.

Table 10 lists the coefficients for quadratic fits to the data: The fit QQ3 is to the three spectroscopic parameters for the sample of 105 stars in 15 clusters. The second quadratic fit, QQA9, uses both the spectroscopic and photometric parameters for 77 stars in 13 clusters. Figures 9 and 10 illustrate the results of the two fits - the quadratic solutions are applied to each star in the two samples and the resulting [Fe/H] values are compared with [Fe/H]$_{ZW}$. These individually calculated [Fe/H] values are then averaged for each cluster and again compared with [Fe/H]$_{ZW}$. The average predicted metallicities and dispersions are listed in Table 11 and illustrated in Figures 11 and 12. The regression lines in Figures 11 and 12 do not quite have unit slope - as do the lines in Figures 9 and 10 - because in the mean plot we simply assigned each cluster unit weight. The differences, however, are small, only a few hundredths of a dex. We can use the scatter shown in Figures 11 and 12 to estimate the uncertainty associated with a typical estimate of [Fe/H] for a cluster which has had 6 to 9 stars in it measured spectroscopically and photometrically. These values are given on the last line of Table 10.

Based on Table 10 and the figures, we conclude the following: First, *all* of the solutions are able to reproduce the [Fe/H]$_{ZW}$ scale with a mean uncertainty of 0.11 dex or less. Second, the addition of the photometric parameters to the solutions reduces the uncertainties from the spectroscopic parameters alone by about 30% for both the linear and quadratic solutions. Third, the two quadratic solutions are a small improvement over their linear counterparts. Table 11 shows that there is a smaller amount of scatter between the [Fe/H]$_{ZW}$ values and the mean cluster values calculated from the quadratic solutions than for the linear ones. Figure 13 shows the close correspondence between the linear and quadratic solutions that use both spectroscopic and photometric parameters for each star. Although one could argue that such a small difference implies that we are unjustified in using a quadratic solution, we would again point to the clearly non-linear relationship between the indices and [Fe/H]$_{ZW}$ (Figures 4, 5, and 6).

### 4.7 The Dependence of the [Fe/H] Calibration on Stellar Luminosity and Other Physical Parameters

Our main goal has been to develop a reliable, accurate method based on near-IR spectroscopy for measuring the metallicity of globular clusters with a minimum of supporting information. In the previous sections we have shown that while the best such measures involve knowledge of color and luminosity, it is still possible to estimate [Fe/H] with an uncertainty of ± 0.10 dex with spectral indices alone. The technique requires the observation of a half dozen or so of the brightest stars in a cluster. The questions we address now are what do we mean by "the brightest stars" and how strongly is the estimate of [Fe/H] affected by which stars are chosen. For example, the dependence of the strength of the near-IR CO absorption bands on stellar luminosity class is well known (*e.g.* Baldwin, Frogel, & Persson 1973). The Na and Ca features also show a significant enhancement in going from dwarfs to giants (*e.g.* Ramírez *et al.* 1997).

For the range in $M_K$ of the globular cluster giants we have observed, the PCA analysis (section 4.4) has shown that the dependence of [Fe/H] on $M_K$ by itself is small. The sense of this small dependence is that at a fixed spectroscopic index strength the estimated value for [Fe/H] declines as the luminosity is increased, i.e. as $M_K$ becomes more negative (see Table 10). Analysis of a data set similar to that presented here for M giants in the Galactic bulge (Ramírez *et al.* 2000a) also shows that over a range in $M_K$ as large as that for the cluster stars there is no significant dependence of EW(Ca) or EW(Na) on luminosity, and that there is only a small dependence of EW(CO) on luminosity.

Origlia & Oliva (2000) demonstrated that observed differences in strength of CO absorption in giants of spectral type mid-K and later with the same [Fe/H] are due primarily to changes in surface gravity and atmospheric microturbulent velocity, *not* on temperature. Ramírez *et al.* (2000b, Fig. 4) shows that these two physical parameters are coupled - higher microturbulence tends to occur in stars of lower surface gravity. Thus, microturbulence increases as one goes up the giant branch. The large effect that small changes in the microturbulent velocity has on CO band strength was first illustrated by Frogel (1971). McWilliam & Lambert (1984) have also demonstrated the importance of microturbulence. The explanation of the effect is simple (Frogel 1971): The observed CO absorption bands are blends of dozens of individual lines, resolvable only at the highest resolutions; most of the lines are saturated. Therefore, even a small increase in velocity broadening due to microturbulence will result in a significant increase in overall absorption.

We will examine two possible situations where the luminosities of the observed cluster giants may effect the outcome of a metallicity determination - first, where only a few of the brightest stars in a cluster can be observed, and second, the case of a poor cluster with only a few bright giants.[7]

Our first test involves the two samples of stars observed in 47 Tucanae mentioned earlier. We obtained spectra for a sample of 7 fainter giants in 47 Tuc in addition to the 7 bright giants used in the [Fe/H] calibration. If we exclude the two that have $M_K$ fainter than −4.0 from the faint sample to maintain our initial selection criteria, the remaining 5 have mean values for $M_K$ and *(J−K)* of −4.7 ±0.3 and 0.90 ±0.03, respectively. These values are 1.5 magnitudes fainter and 0.2 magnitudes bluer than the calibration sample (Table 4). If we apply the quadratic solution based only on the spectroscopic indices (QQ3, Table 10) to these stars we get a mean [Fe/H] of −0.80 ±0.07 compared with −0.72 ±0.12 for the bright sample used for the calibration. The same exercise with the quadratic solution based on both indices and photometric parameters yields −0.63 ±0.03 and −0.75 ±0.10 for the faint and bright calibrating samples, respectively.

For the second test we considered just the three brightest stars observed in each of the calibrating clusters (only relative magnitudes are important here), applied one of the solutions to them, and compared the results obtained with the results for the full sample of stars. This test has the added benefit of assessing the effect of observing such a small number of stars in a cluster as opposed to the 6 to 9 typically observed. The mean of the differences in the sense (full sample) – (3 brightest stars sample) for the quadratic solutions based first on just spectroscopic indices and then on indices plus photometric parameters are −0.02 ±0.05 and +0.01 ±0.04, respectively. Furthermore, even with only 3 stars per cluster the means for the *intra*-cluster dispersions are only 0.01 dex greater for the two solutions than for the full samples.

As a final test, we observed 4 bright stars near the center of NGC362 with no previous photometric observations of any kind. These are numbered 1 through 4 in Table 2. Application of the quadratic solution QQ3 from Table 10 to the spectral indices for these 4 stars yields a mean [Fe/H] of −1.08 ±0.21. This should be compared with [Fe/H] = −1.16 ±0.19, the average value for solution QQ3 applied to the calibrating stars in this cluster.

We conclude that our technique for calculating [Fe/H] for globular clusters is robust against small sample size and a range in the luminosities of the stars in the sample provided that the sample should not include stars fainter than $M_K \approx −4.0$. Since this is between two and three magnitudes fainter than the tip of the RGB for nearly all clusters, it does appear to be an important restriction.

## 5. ON THE USE OF EW(CO) ALONE TO DETERMINE [FE/H]

It has been known for some time that the strength of the first overtone CO absorption band in globular cluster giants is sensitive to the [Fe/H] of the cluster (Frogel *et al.* 1978; Aaronson *et al.* 1978; Frogel *et al.* 1983a). This is true for both observations of CO in the integrated light of the cluster and for the strength of the feature in individual stars of the same luminosity. Saturation effects aside for the moment, we expect that [Fe/H] and EW(CO) should scale approximately linearly, not quadratically as might be expected for a diatomic molecule. The reason is that for K or M giants with [O] > [C], essentially all of the carbon is locked up in CO. There is no free C and any other carbon bearing molecules have partial pressures that are at least several orders of magnitude lower than that of CO (*e.g.* Tsuji 1964). Hence, if other physical parameters are held constant, the CO absorption strength will scale linearly with [C/H].

---

[7] In connection with the latter case we point out that Frogel *et al.* (1983a) showed that over a considerable range in cluster properties, even the least populous clusters in their sample had at least one star within a couple of tenths of a magnitude of the theoretical upper luminosity limit for the RGB.

Since the stars we have observed in each cluster all lie close to the tops of their respective RGBs, and the RGBs shift to systematically cooler temperatures with increasing [Fe/H], we cannot compare stars at constant $T_{eff}$ for the full range of cluster [Fe/H] values encountered. Origlia & Oliva (2000) point out that the correlation between CO strength and [Fe/H] at constant luminosity arises not only because of metallicity variations but because of this systematic shift of the RGB with metallicity. Furthermore, as we have shown earlier, in spite of the fact that EW(CO) is many times stronger than EW(Na), the latter carries more weight in our multi-parameter solutions because the relative star to star scatter within a cluster for the Na I feature is considerably less than it is for the CO absorption.[8] Nevertheless, because the CO absorption is nearly an order of magnitude greater than that of any other feature in the near IR for cool stars, it is worthwhile to re-examine the efficacy of EW(CO) by itself in estimating [Fe/H] for globular clusters.

Table 10 lists the parameters for the linear (LL1) and quadratic (QQ1) solutions to [Fe/H] as a function of EW(CO) alone for the sample of 105 stars defined earlier. Both are clearly inferior to their counterparts based all three spectroscopic indices. The correlation coefficients are less, though still significant, and the estimated dispersions are about twice as large. Inspection of Table 11 reveals that the EW(CO) will systematically underestimate [Fe/H] for the most metal rich clusters (solution QQ1 - we did not tabulate results from LL1 since they are nearly identical to those from QQ1). This result is not surprising given the appearance of Fig. 4. The second overtone CO absorption at 1.62 μm in the H-band as a metallicity indicator gives a similar result (cf. the left hand side of Fig. 5 in Origlia *et al.* 1997), i.e. high "true" [Fe/H] values are underestimated.

Why does EW(CO) - and other measures of CO absorption strength - become insensitive to [Fe/H] for very metal rich stars? As we discussed in section 4.7, for cool giants the microturbulent velocity plays a major role in determining the observed band strength because the individual lines that make up the bands are nearly all saturated. Broadening due to microturbulence will increase the absorption in the wings of each line thereby strengthening the band. We conclude, therefore, that the loss of sensitivity of EW(CO) to [Fe/H] at the high metallicity end of the globular cluster distribution is probably due to saturation effects that are not offset by an increasing value of the microturbulent velocity. In spite of this limitation, though, the CO absorption by itself can be a useful estimator of [Fe/H], especially for clusters with [Fe/H] ≤ -0.6.

## 6. FURTHER DISCUSSION AND CONCLUSIONS

It has been clear for quite some time that the metal rich globular clusters of the Milky Way form a separate subsystem of clusters distinct from those in the halo of the Galaxy and probably associated with the Galactic bulge. A study of these metal rich globulars, then, has the potential to help us understand the formation and early chemical evolution of the central parts of the Milky Way provided that their metallicity can be established with a reasonable degree of accuracy. While photometric indices in both the optical and near-IR can be used for [Fe/H] estimates, spectroscopy is the technique of choice for obtaining the most accurate values for [Fe/H]. Because most bulge globulars are heavily reddened and are located in crowded fields, infrared spectroscopy is expected to provide better results than optical spectroscopy.

---

[8] As mentioned earlier, a likely explanation for the large intra-cluster scatter in EW(CO) is star-to-star differences in mixing history (see Kraft 1994 for a review) . This certainly seems to be the case for 47 Tucanae. The large star-to-star variations observed in the photometric CO indices for the 47 Tucanae giants (Frogel *et al.*1981) are closely anti-correlated with optical CN band strength variations.

In this paper we describe a method for obtaining [Fe/H] values for globular clusters with moderate resolution K band spectroscopy of about half a dozen of the brightest RGB stars in a cluster. We determine three spectroscopic indices: EW(Na, Ca, CO). These three indices are highly correlated with one another and with [Fe/H] but only weakly with $M_K$. The observed intra-cluster dispersions in EW(Ca) and EW(Na) are consistent with measurement uncertainties while the dispersion in EW(CO) is significantly larger. We attribute this latter result to differences in the mixing histories of RGB stars as they evolve up the RGB.

We calculate both linear and quadratic fits of [Fe/H]$_{ZW}$ to the indices for different sets of parameters and find that the quadratic fit gives a somewhat better than the linear one. The resulting values for [Fe/H] are able to reproduce the abundance scale of Harris (1996), which is basically the scale of Zinn & West (1984), with an accuracy of about ±0.10 dex over the range −1.8 < [Fe/H] < 0.0. Although we used 6 to 9 stars per cluster to derive the abundance scale, as few as three giants in a cluster will yield an [Fe/H] value nearly as good as the larger sample. Although the near-IR spectroscopic scale does not require knowledge of a clusters' distance or reddening or any photometric parameters for the stars, we show that if you do know $M_K$ and $(J-K)_0$ of the stars for which spectra are available, then the predicted uncertainty in the average derived value of [Fe/H] for a cluster decrease by several hundredths of a dex. The parameters for all of the fits are given in Table 10, while a comparison with the Harris/Zinn & West values is in Table 11.

In a subsequent paper (Frogel, Stephens, & Ramírez 2001) we will present near-IR spectroscopy of giants in a number of poorly studied bulge clusters and use the present calibration to derive accurate metallicity estimates for them.

## 6.1 Additional Considerations of the CO Absorption

The first overtone band of CO absorption at 2.29 μm - which in the past has most often been measured photometrically with two narrow band filters - is frequently used as a metallicity indicator. Because of its great strength, observations with NGST may rely on measurements of the CO band for metallicity estimates of distant star clusters and galaxies. First, we show that for globular cluster giants this photometric color index is tightly correlated with EW(CO) measured spectroscopically. Houdashelt (1995) had previously shown that the photometric CO index measured for early-type galaxies is also well correlated with spectroscopic measures of its strength. We then demonstrate that for globular stars near the top of the RGB, EW(CO) appears to lose sensitivity to [Fe/H] for [Fe/H] values greater than about −0.4, most likely due to saturation effects. This limits the usefulness of EW(CO) by itself for determining [Fe/H] in *individual stars* of higher metallicities. On the other hand, if you measure EW(CO) in the *integrated light* of a stellar system, the sensitivity at the high end of the [Fe/H] scale will be regained. The reason for this is because even though the band is saturated in the coolest, most luminous giants, it does not appear to be saturated in stars further down the RGB. As the whole RGB shifts to cooler temperatures as [Fe/H] increases, integrated over the entire RGB the observed CO strength will steadily increase. This was demonstrated by Aaronson *et al.* (1978) who measured photometric CO indices in the integrated light of several dozen globular clusters. Their measured CO indices are plotted in Fig. 14 along with the revised ZW abundances of Harris (1996). The least squares solution is given by equation 5. The R value is 0.84

$$[Fe/H]_{ZW} = 13.94 \times CO - 2.01 \qquad 5$$

The fact that thermal emission from the atmosphere and telescope is the dominant source of noise longward of about 1.9 μm for observations from the ground makes it difficult to measure the strength of the CO band at 2.29 μm in faint objects. Furthermore, because it is close to the red atmospheric cutoff (due to $H_2O$) of the K band, galaxies with redshifts greater than about $z$=0.02 (the Coma cluster), are impossible to measure from the ground. Observations from space, on the other hand, get around these difficulties. With NGST, for example, it should be possible to measure the CO band strength in globular clusters in the Coma cluster and beyond. Therefore, although [Fe/H] estimates with CO by itself are not as accurate as those that result from also measuring the Na and Ca features, CO measurements may be the only way of determining the metallicity distributions of distant globular cluster systems. In addition, narrow band imaging with an appropriate set of filters could be used to map out the CO strength in distant galaxies with NGST and be a great aid in studying their stellar populations.

### 6.2 What Abundance are We Really Measuring?

Finally, we consider the issue raised by RHS97 and by Carretta & Gratton (1997), *viz.* that there may be a systematic difference between the Zinn & West (1984) scale and that based on high resolution spectroscopy, especially at the metal rich end. This is an important point because our new IR-based metallicity scale is calibrated on the Harris revision of the ZW scale. Carretta & Gratton's comparison of high resolution spectroscopic data to the Zinn & West scale (their Fig. 5) leads them to the conclusion that "the ZW scale is far from linear," and that the ZW scale yields [Fe/H] values that are too high, on average, by about 0.1 dex for [Fe/H] > − 1.0. The non-linear fit they present in their Fig. 5 suggests that the spectroscopic abundances diverge more and more from the ZW scale as [Fe/H] increases. RHS employ the W' calcium triplet index as a surrogate for [Fe/H] and find a non-linearity in the same sense as found by Carretta & Gratton, *viz.* that [Fe/H] on the ZW scale increases more rapidly at the metal rich end than one would expect from the increase in W'. For both of these studies there is a paucity of good abundance determinations at the metal rich end of the scale as we discuss in the Introduction to this paper. Although we have shown that a quadratic fit of the infrared indices to the ZW scale is to be preferred, the difference between it and a linear fit is small.

For the above reasons we present here a brief comparison of the ZW scale as updated by Harris (1996) and the high resolution spectroscopic scale with additional values at the critical metal rich end from Cohen *et al.* (1999) and Carretta *et al.* (2001). In Carretta *et al.* [Fe/H] for NGC 6528 is given as +0.07 based on Keck HIRES data. These authors, with better data in hand, revise the [Fe/H] value given earlier by Cohen *et al.* for NGC 6553 upwards to +0.04. If we add these two values to the mean [Fe/H] values from high resolution spectroscopic data in column 4 of Table 8 in CG97 and substitute Harris' (1996) updated values for the ZW ones, we find that the quadratic fit to the two sets of data is nearly indistinguishable from a simple linear fit. Both have R values of 0.99. The only difference is in the zero point. The two equations are:

$$[Fe/H]_{CG} = 0.144 + 0.999[Fe/H]_{ZW}$$

and (6)

$$[Fe/H]_{CG} = 0.210 + 1.132[Fe/H]_{ZW} + 0.053[Fe/H]_{ZW}^2$$

Figure 15 illustrates the relation between the two abundance scales and the linear fit between them.

We conclude, then, that our original objectives have been met. We have set up a near-IR based abundance scale for globular clusters that is easily reproduced. To use it requires medium resolution K-band spectra of a few of the brightest giants in a cluster. Our scale can reproduce the ZW abundance scale as updated by Harris (1996) to ±0.10 dex or better from the metal rich end of the globular cluster distribution down to an $[Fe/H]_{ZW}$ of −1.7. Extension of our calibration to the integrated light of clusters will allow its application to systems well beyond the local group. In fact, it should be possible to apply a stripped down version of our technique that uses only the strong CO band to globular clusters in the Coma cluster of galaxies with observations from the NGST, albeit with reduced accuracy. Finally, again with additional calibration, our method can be a valuable adjunct in the study of stellar populations of galaxies.

Many members of the CTIO mountain staff helped to ensure that our observing runs would be carried out successfully. We thank Ron Probst and Bob Blum for considerable assistance in getting the IRS and OSIRIS up and running with the tip tilt secondary on the Blanco 4-meter telescope. OSIRIS was built with funds from NSF grants AST 90-16112 and AST 92-18449. Richard Pogge supplied us with `blint`, a program used to measure line strengths. JAF and AWS thank Pascale Jablonka for arranging an extended visit at the Observatoire de Paris, Meudon during which time parts of this paper were prepared. We thank Judy Cohen for communicating the results of her work with Gratton *et al.* in advance of publication and for several helpful conversations. The beginning of this long term project - but not the middle or end - was supported by the NSF under grant AST-9218281 to JAF. Instead, we have received considerable assistance from the OSU Astronomy Department and the American Airlines Frequent Flyer Program for our trips to CTIO. We are particularly grateful for assistance from Malcolm Smith, CTIO director.

APPENDIX: ADDITIONAL NOTES ON INDIVIDUAL CLUSTERS

In this Appendix we comment on other recent observations of relevance for the clusters we have observed.

1. NGC 288 & NGC 362

Shetrone and Keane (2000) report the results of a spectroscopic study of about one dozen giants in each of NGC 288 & 362. They derive *overall* metal abundances for these clusters of −1.39 and −1.33, respectively, with quoted uncertainties of only ±0.01 dex. Both of these new determinations of [Fe/H] for the clusters are less than the values listed by Harris of − 1.24 and − 1.16. The fit of our spectroscopic data for these clusters with the Harris values in our calibration is excellent as can be seen from Table 11 and Figures 11 and 12. If we had used the Shetrone & Keane values the scatter in the calibration would have increased.

## 2. NGC 5927

Cohen & Sleeper (1995) obtained new near-IR photometry for 4 clusters in our sample: NGC 5927, 6528, 6553, and 6624. While their ordering of the [Fe/H] values for these clusters agrees with that of Harris (1996), there are differences in detail. For NGC 6528 and 6553, the two most metal rich clusters (according to Harris) in our sample, Cohen & Sleeper derive abundances that are about 0.3 dex greater than those of Harris. Our spectroscopic values are in agreement with Harris'. For NGC 5927 and 6624 they derive values lower than those of Harris. We have already noted that our data for NGC 6624 suggest that the Harris (1996) value is too high as well. Our data (Table 11) would be more consistent with a value of about – 0.60, quite close to Cohen and Sleeper's (1995). This agreement could be fortuitous since we are in closer agreement with Harris (1996) for the other three clusters.

## 3. NGC 6304

Ortolani, *et al.* (2000) present BV photometry for this cluster. Based on its CMD and the location of its RGB, Ortolani *et al.* estimate that NGC6304 has an [Fe/H] value intermediate to those of NGC 6528 and 47 Tuc, or [Fe/H] = –0.57 with $[Z/Z_\odot]$ = 0.05. Their [Fe/H], then, is consistent with Harris' (1999) value of – 0.59 but somewhat less than ours which is about 0.1 dex greater (see Table 11).

## 4. NGC 6440

Minniti (1995) measured the strength of Mg 5175 and Fe 5270 and 5335 in bright globular cluster giants chosen on the basis of their position in a *K*, *J–K* CMD. From these data he finds [Fe/H] = –0.50 for NGC 6440, the same as that given by Harris within the uncertainties of the determinations.

## 5. NGC 6528

Ortolani, Barbuy, & Bica (1991) derived a *relative* [Fe/H] ranking based on the curvature of a cluster's RGB in a *V*, *V–I* CMD, an idea first proposed by Lloyd Evans & Menzies (1977). Ortolani *et al.* found the following ranking, in order of increasing metallicity: 47 Tucanae, NGC 6528, and NGC 6553. They also concluded that the RGB curvature for NGC 6553 was the same as for field stars in Baade's Window and therefore the cluster must have a similar metallicity to the field stars. This implies an [Fe/H] of about –0.2 based on recent estimates of this quantity for Baade's Window.

Bica *et al*. (1998) determine $[Z/Z_\odot]$ for a number of metal rich clusters based on the CaII triplet. Because of α element enhancement their Z values are expected to be greater than true [Fe/H] determinations. They find +0.2 for NGC 6528 and – 0.3 for NGC 6624. On their scale, the $[Z/Z_\odot]$ for 47 Tuc is –0.6. These values are all higher than the [Fe/H] values give by Harris (1998), the large positive value for NGC6528 is especially so.

Cohen & Sleeper's (1995) also observed this cluster as we discuss under NGC 5927 earlier in this Appendix.

## 6. NGC 6553

Guarnieri *et al.* (1998) have *JK* and *VI* data for stars in NGC 6553. Based on the slope of the RGB method (Kuchinski & Frogel 1995) they find [Fe/H] = −0.18 ±0.17, while from the *V − I* color of the RGB at the level of the HB, they find [Fe/H] = −0.28 ±0.15, in close agreement with the former value. Both values are in close agreement with the Harris (1999) value.

See Cohen & Sleeper's (1995) discussion of this cluster under NGC 5927.
See the reference to Ortolani *et al.* (1991) under NGC 6528.

## 7. NGC 6624

See Cohen & Sleeper's (1995) discussion of this cluster under NGC 5927.

## 8. M69

Ferraro *et al.* obtained *BVJK* photometry of RGB and HB stars in M69. They find a strong similarity to 47 Tuc and, from optical and near-IR CMDs, derive [Fe/H] = −0.75 ±0.20 for M69. Our spectra indicate a metallicity slightly higher than that of 47 Tuc (Table 11). Inspection of Table 4 shows that the mean value for each of the 3 spectroscopic indices for the M69 stars is about 1 $\sigma$ greater than the corresponding average for 47 Tuc even though the average color of the M69 stars is a bit bluer. The average magnitudes of the two clusters are closely similar. Our measurements of EW(Na) and EW(Ca) for the M69 stars may be affected by fringing on the detector (see Fig. 3)

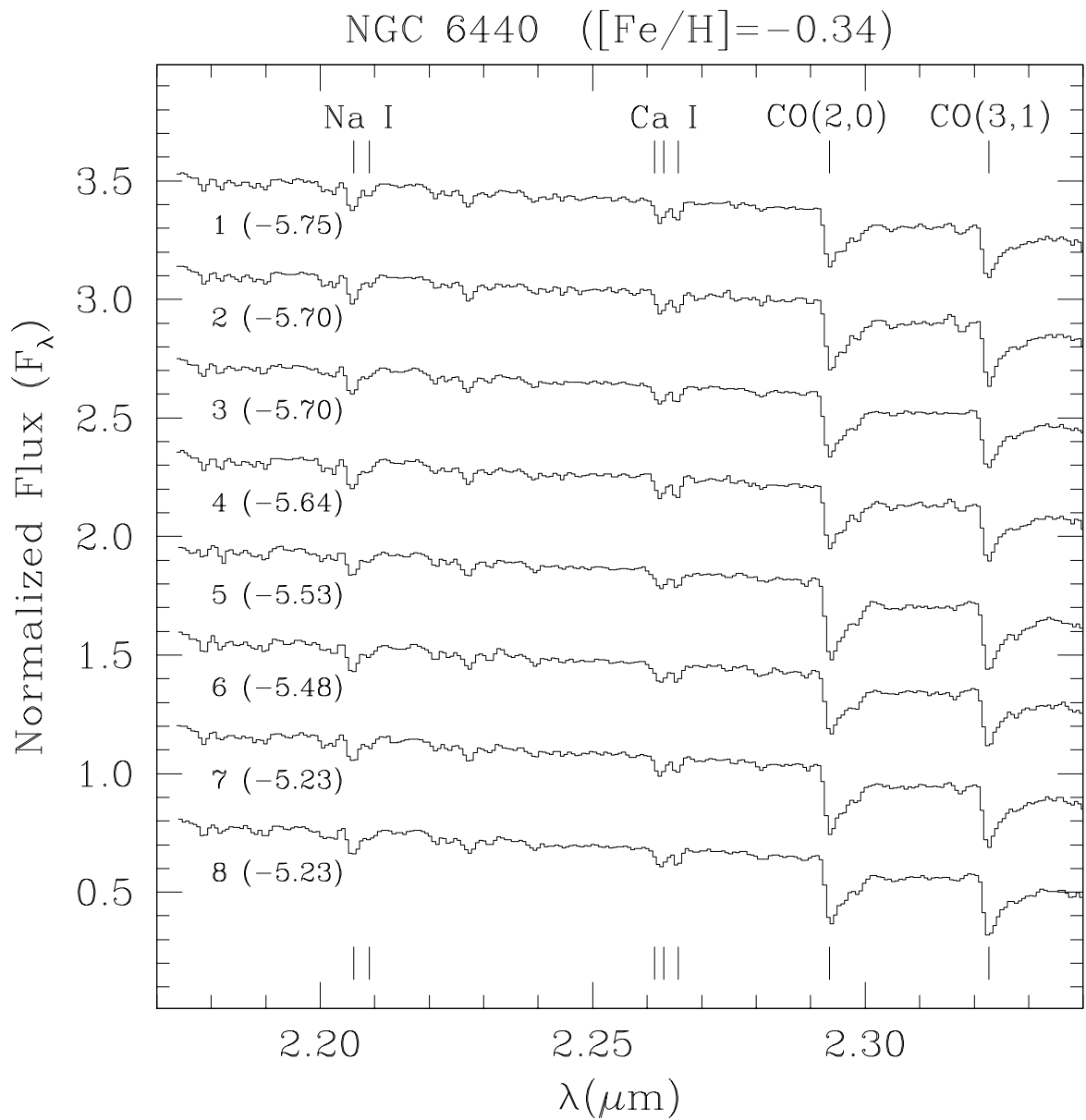

Fig. 1 — The K-band spectra of the 8 stars in NGC6440, a bulge globular cluster with [Fe/H] = -0.34 (Harris 1996). The absolute *K* magnitude for each star is given in parentheses. The position of the Na doublet, the Ca triplet, and the first two band heads of CO are indicated. Note the close similarity of all of the spectra; features significantly weaker than the ones we have measured are clearly visible on all 8 of the spectra.

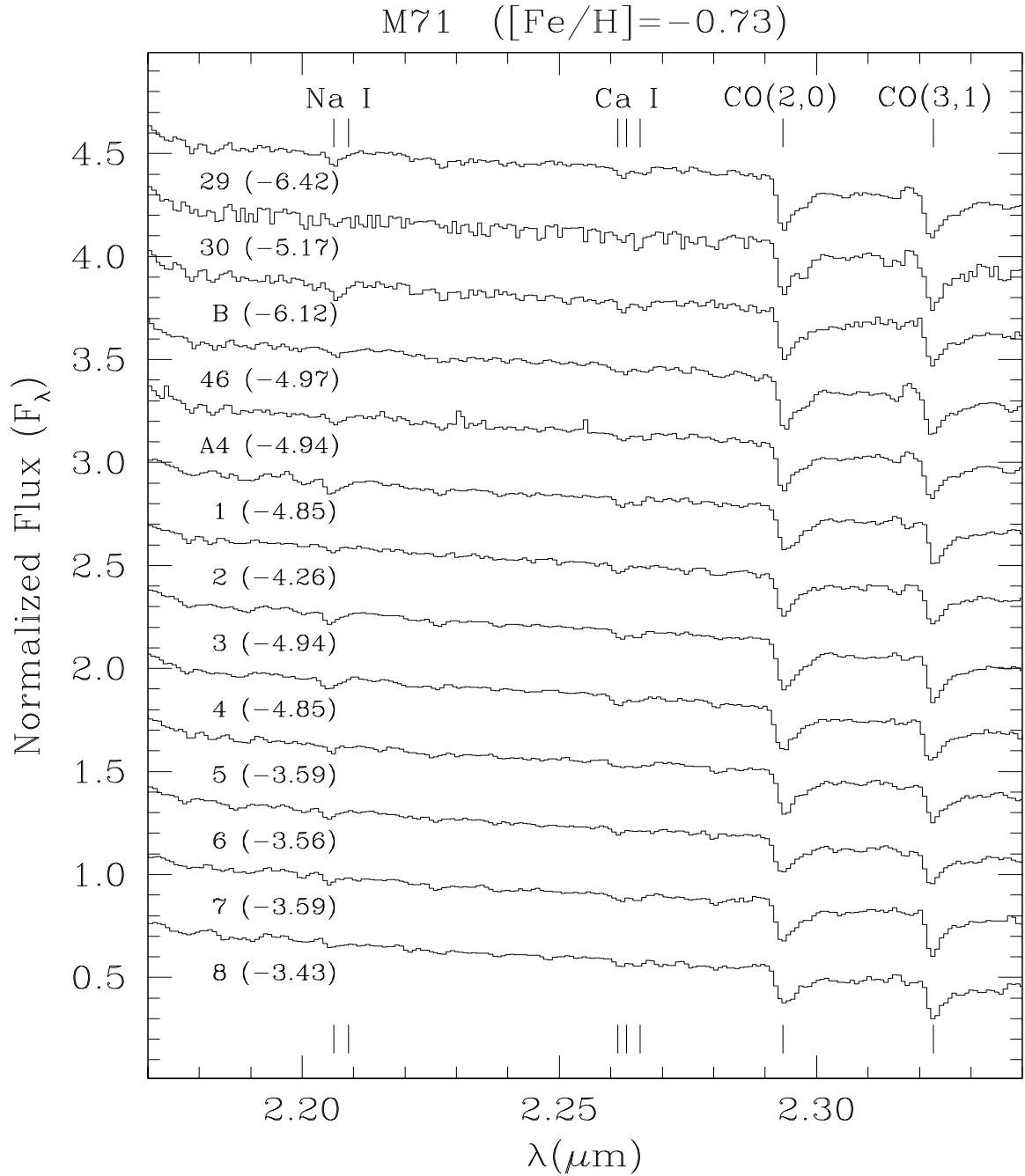

Fig. 2 — This figure is similar to 1 except that it displays the observations of the stars in M71 (NGC 6838) with an [Fe/H] = -0.73. The range in $M_K$ is much greater than for NGC6440 and the spectra for a few of the stars appear to be somewhat noisier than the rest, especially M71-30.

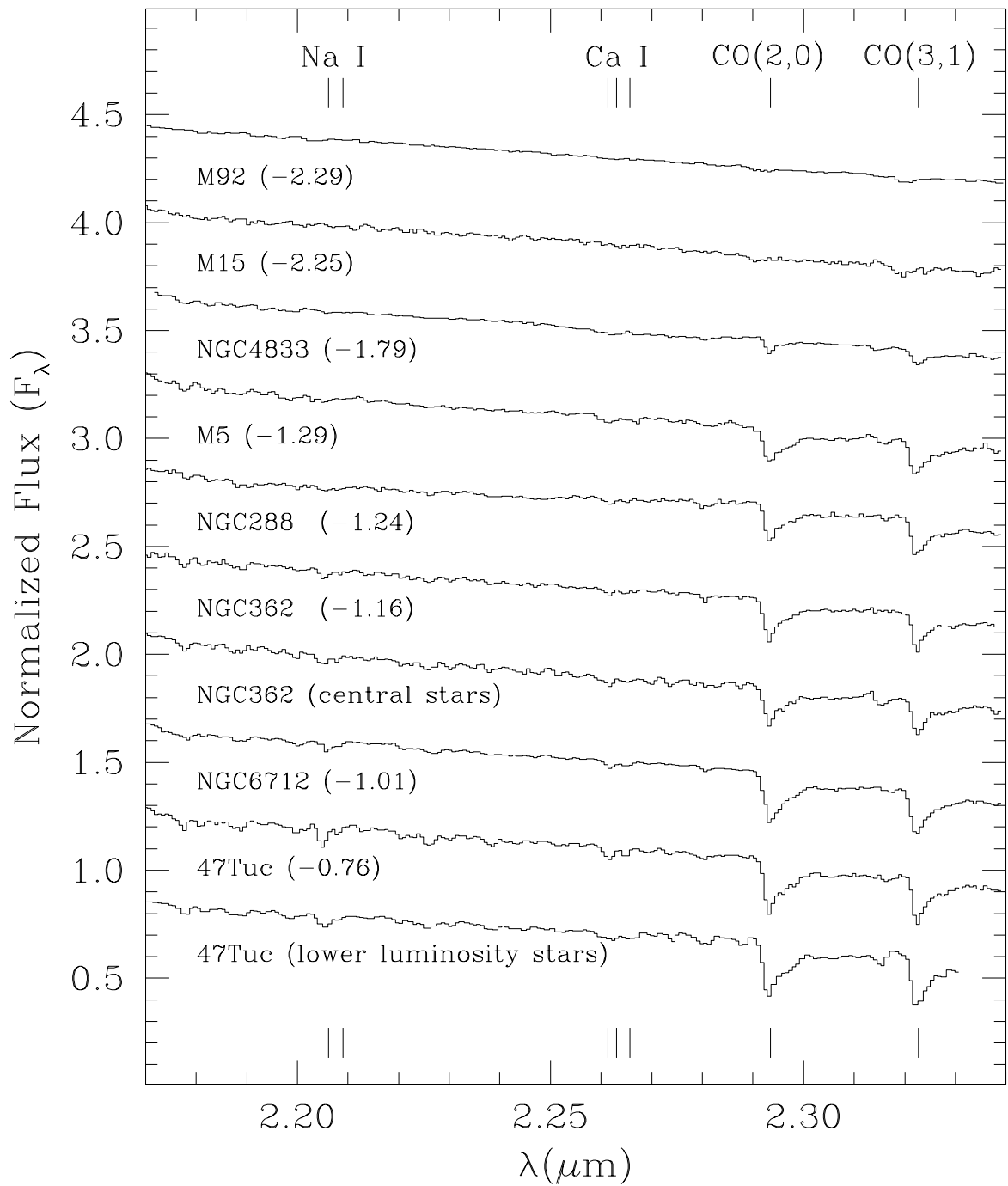

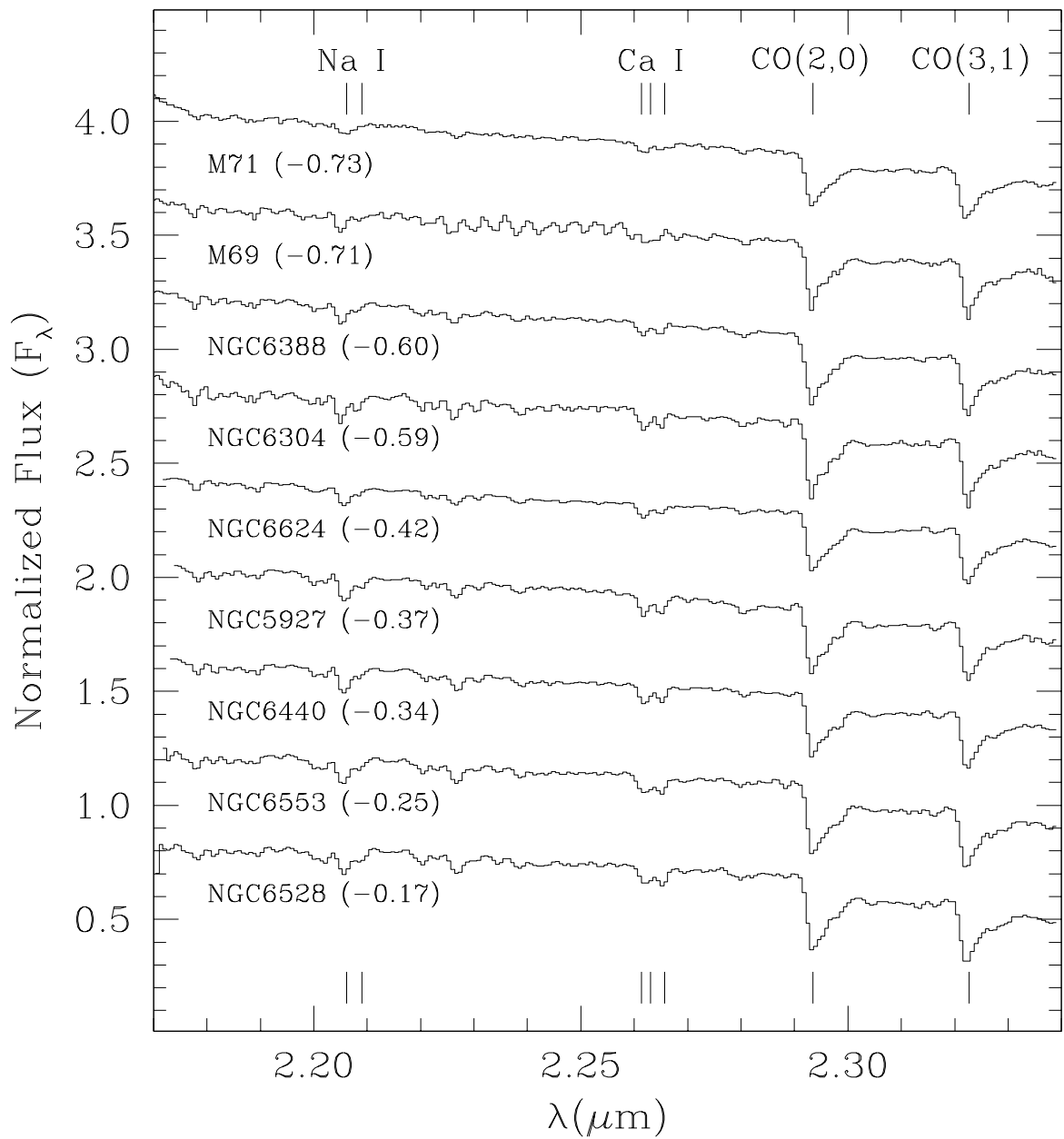

Fig. 3 — The two parts of this figure display the average of the individual stellar spectra for each cluster observed. Note that the spectra for the stars in M69 are affected by fringing on the detector. Also note the near featurelessness of the spectra for M15 and M92. The metallicity of each cluster is given in parentheses.

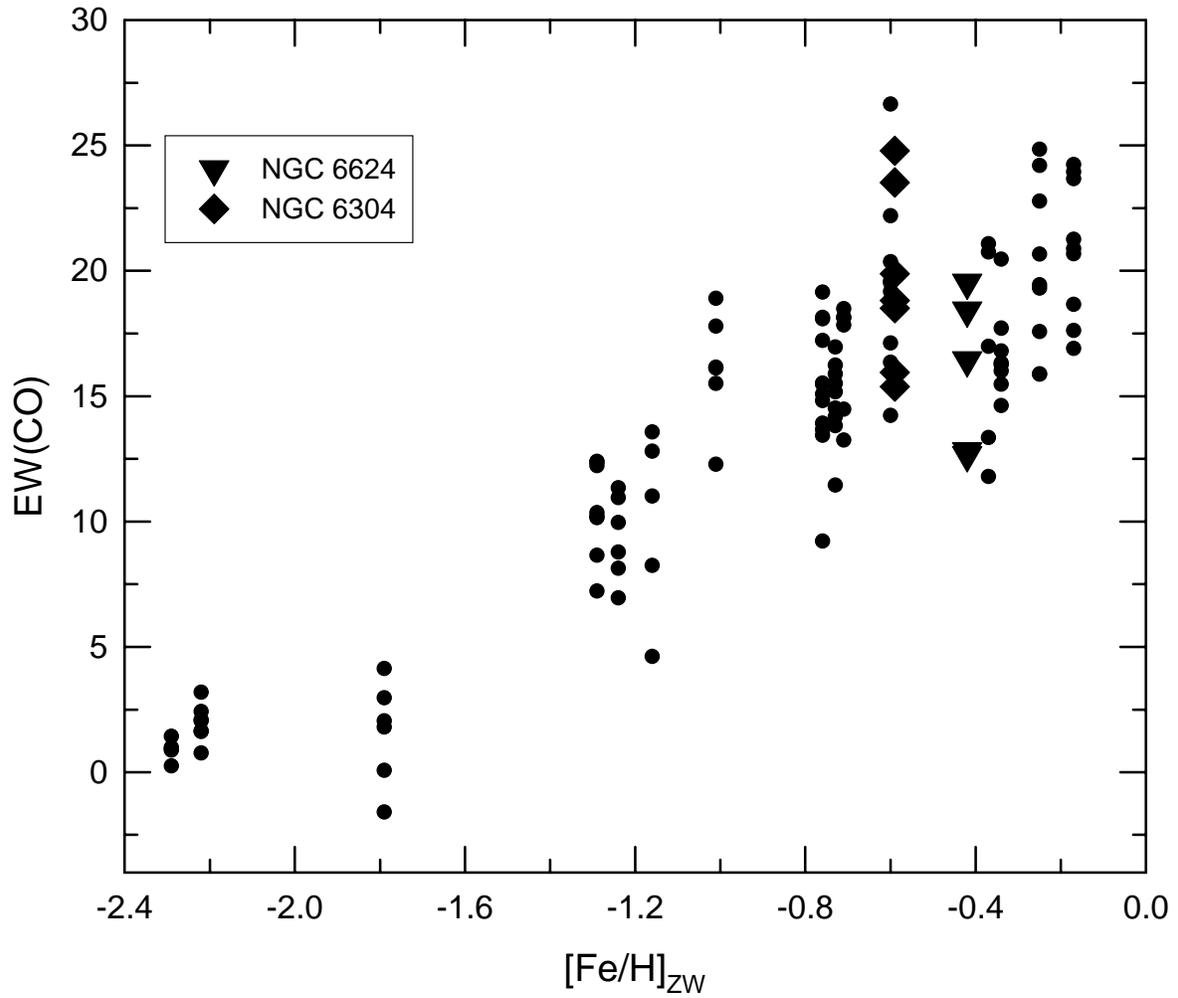

Fig. 4. — The spectroscopically measured value of EW(CO) for the 2.29 μm CO absorption band for each observed star is plotted versus cluster abundance on the Zinn & West scale as revised by Harris (1996). This scale is referred to as [Fe/H]$_{ZW}$ in this paper. The two clusters whose stars are plotted with distinct symbols - NGC 6304 and 6624 - are discussed in the text.

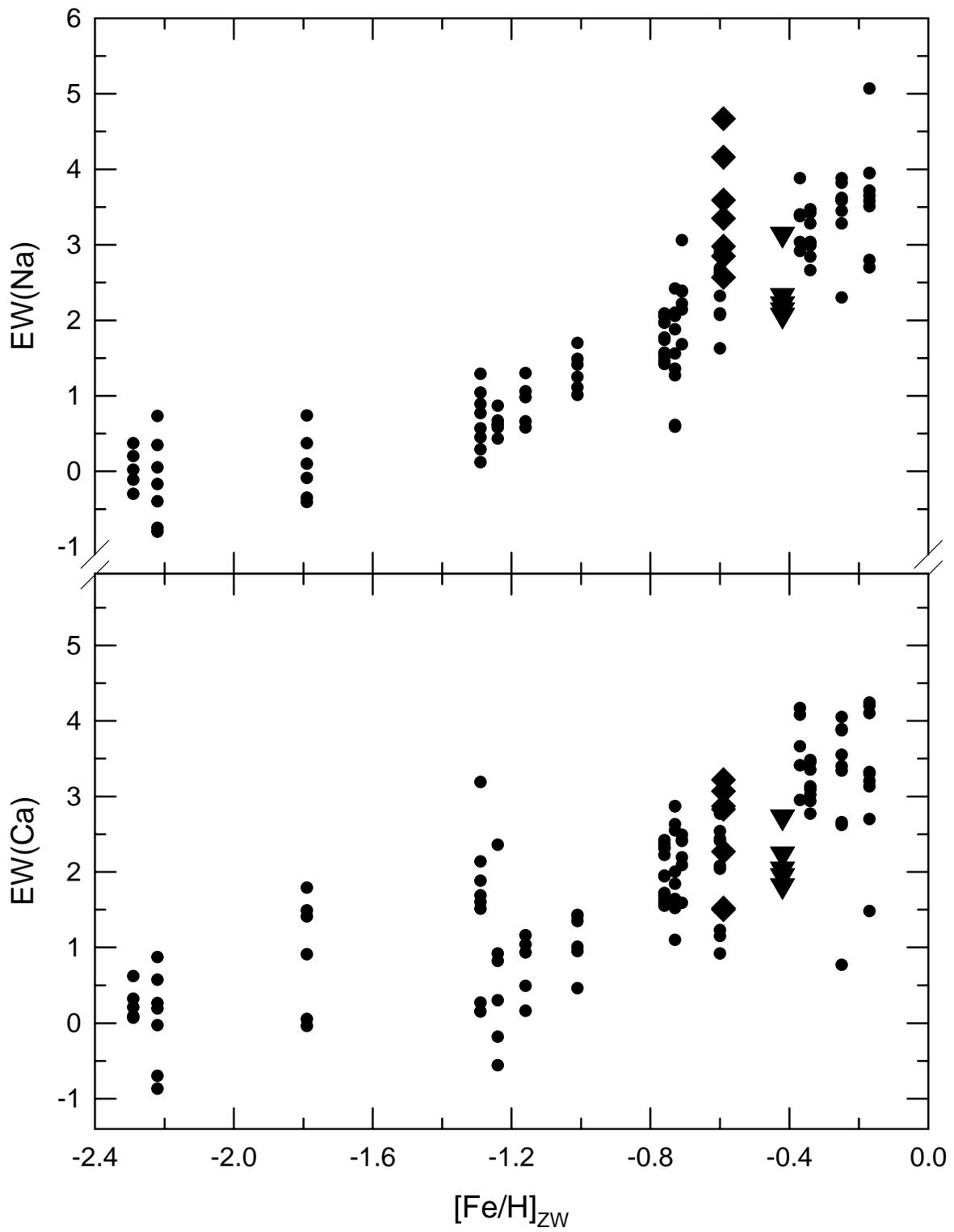

Fig. 5. — The same as Fig. 4 except that these plots show EW(Na) and EW(Ca) versus cluster abundance on the [Fe/H]$_{ZW}$ scale.

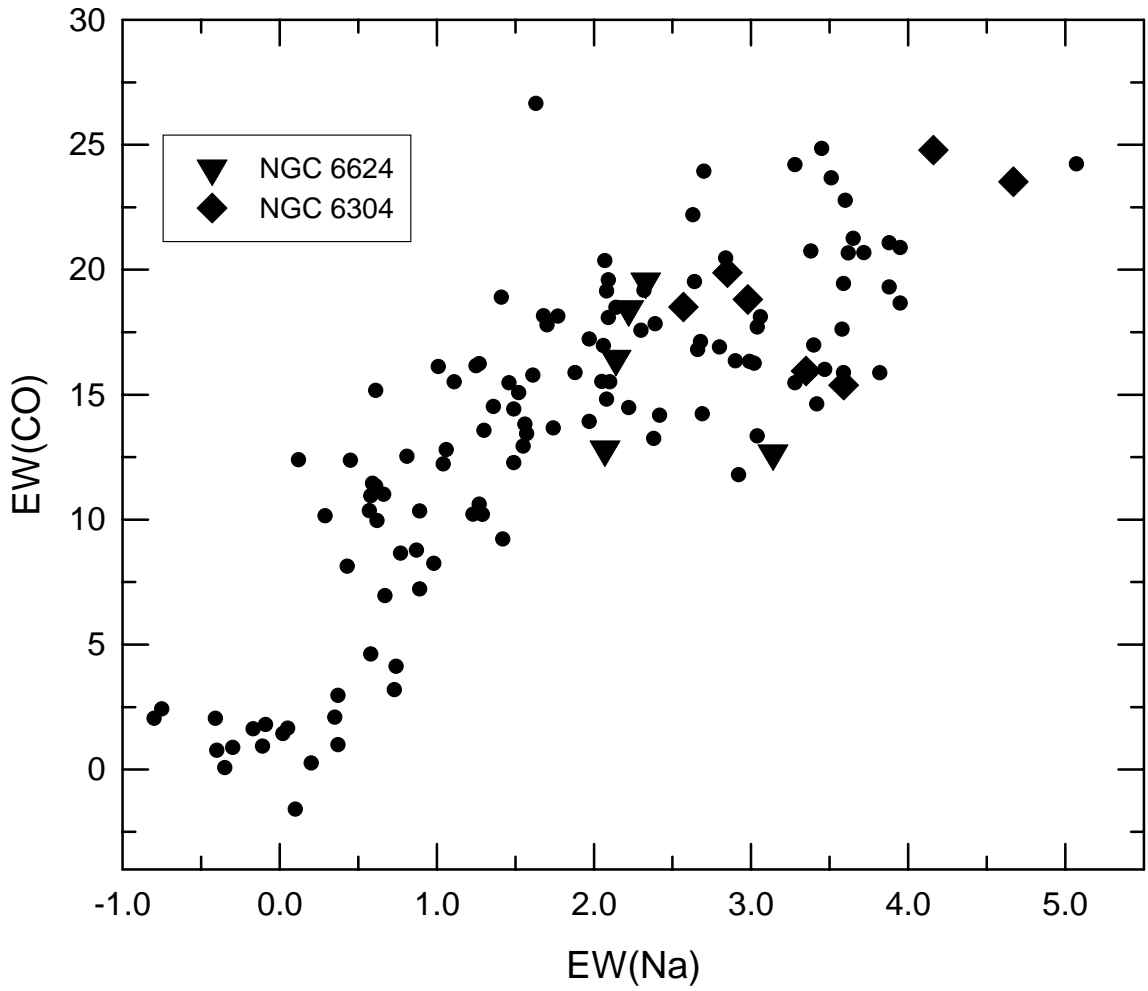

Fig. 6. — EW(CO) is shown as a function of EW(Na) for all stars in the spectroscopic sample. Note the non-linear nature of this dependence. Also note that stars from NGC 6304 and 6624 follow the same relationship as stars from the other clusters.

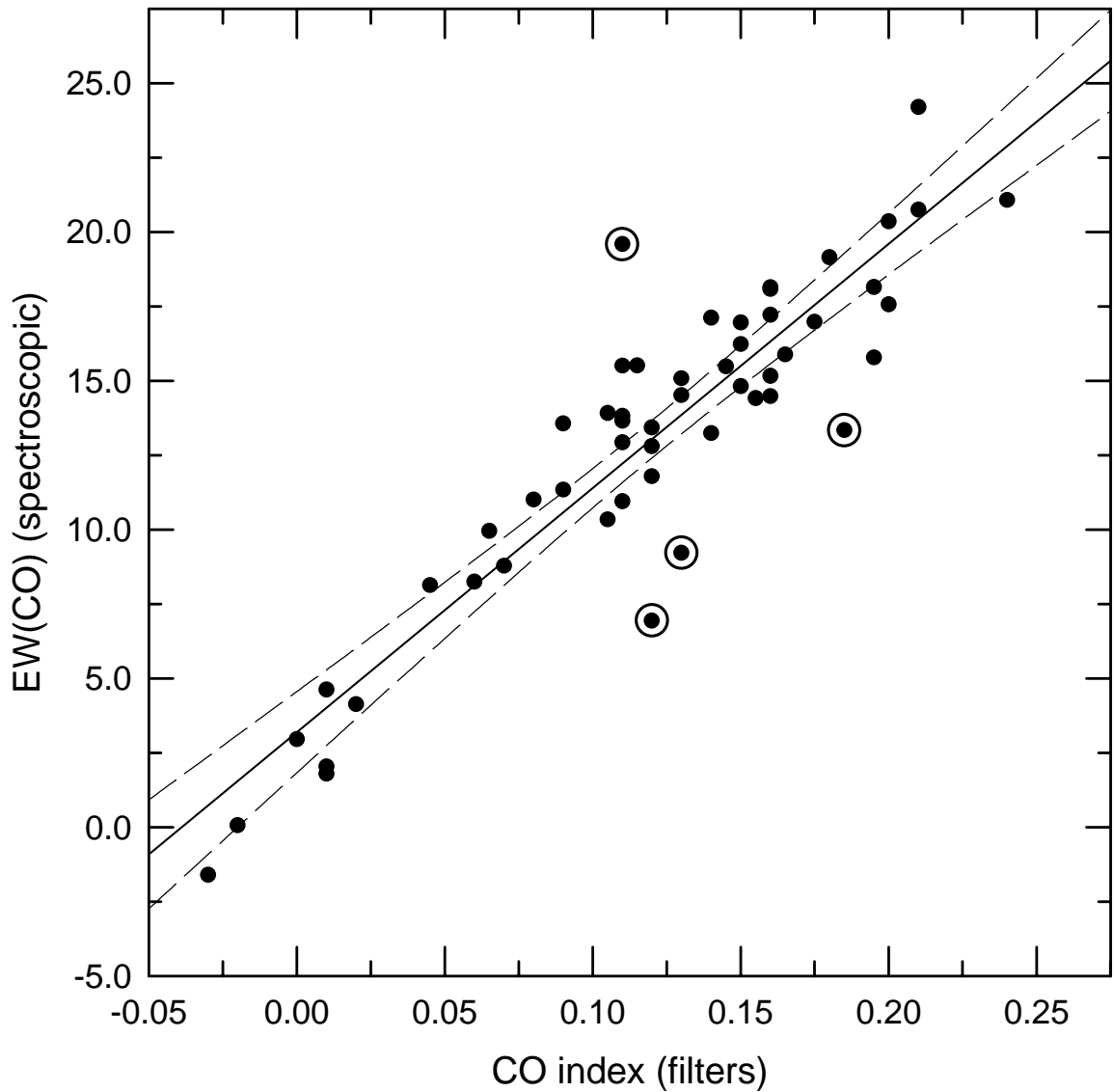

Fig. 7. — The relationship between the spectroscopically determined EW(CO) and the photometric CO index (Frogel, *et al.* 1979, 1981, 1983b; Frogel 1985) is illustrated. The filters which define the photometric index are described in Frogel *et al.* (1978). The solid line shows the best fit least-squares relation between the two quantities. The dashed lines are the 95% confidence limits.

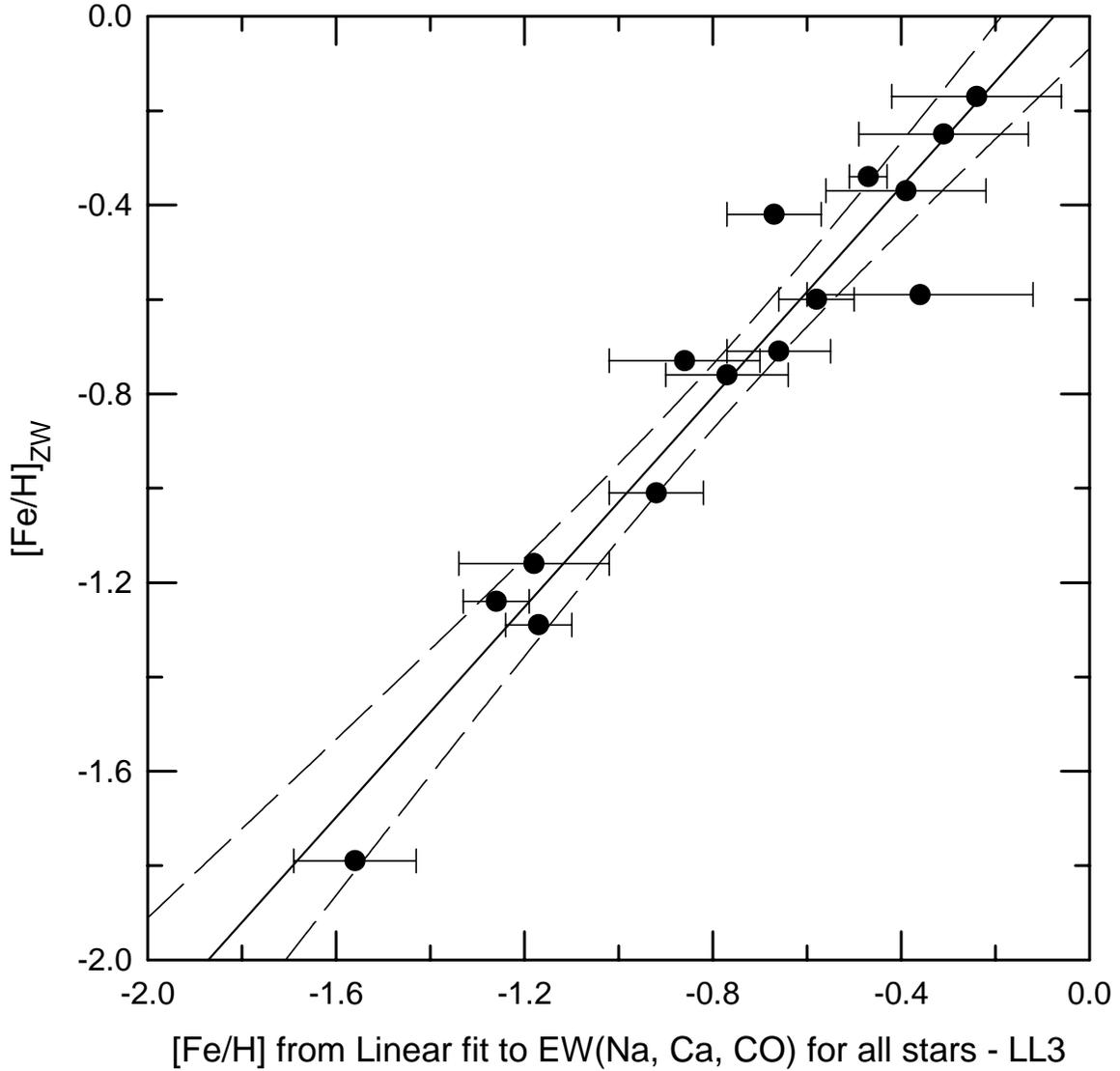

Fig. 8. — A linear least squares fit of $[Fe/H]_{ZW}$ to the three spectroscopic indices for each of the 105 stars from the 15 calibrating clusters was found (Table 10, solution LL3). With the resulting equation (eqn. 3) a value of [Fe/H] was calculated for each star from the spectroscopic indices and then an average [Fe/H] was found for each cluster. These averages are the values on the horizontal axis. The horizontal bars point indicate the dispersion about the mean for each cluster's stars. The straight line shows the best fit relation between these mean values and $[Fe/H]_{ZW}$ values. Also shown are the 95% confidence limits.

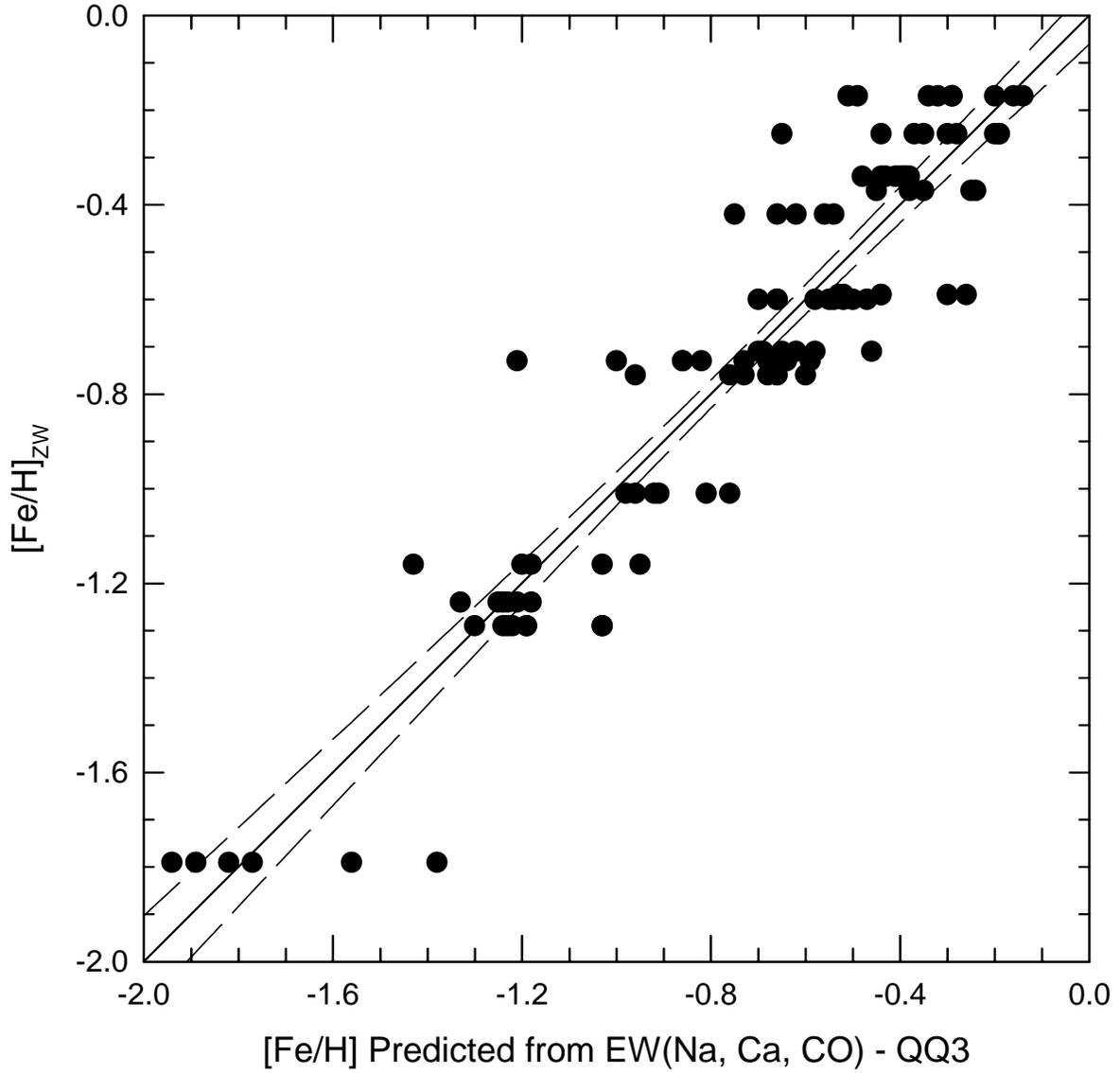

Fig. 9. — We solved for a quadratic fit between [Fe/H]$_{ZW}$ for each cluster and the three spectroscopic indices for the 105 stars in the spectroscopic sample (Table 10, solution QQ3). Then we applied the fit to each star to calculate individual [Fe/H] values. These are the values on the horizontal axis of this figure. The dashed lines in the figure are the 95% confidence limits about the best fit line which is the line of equality.

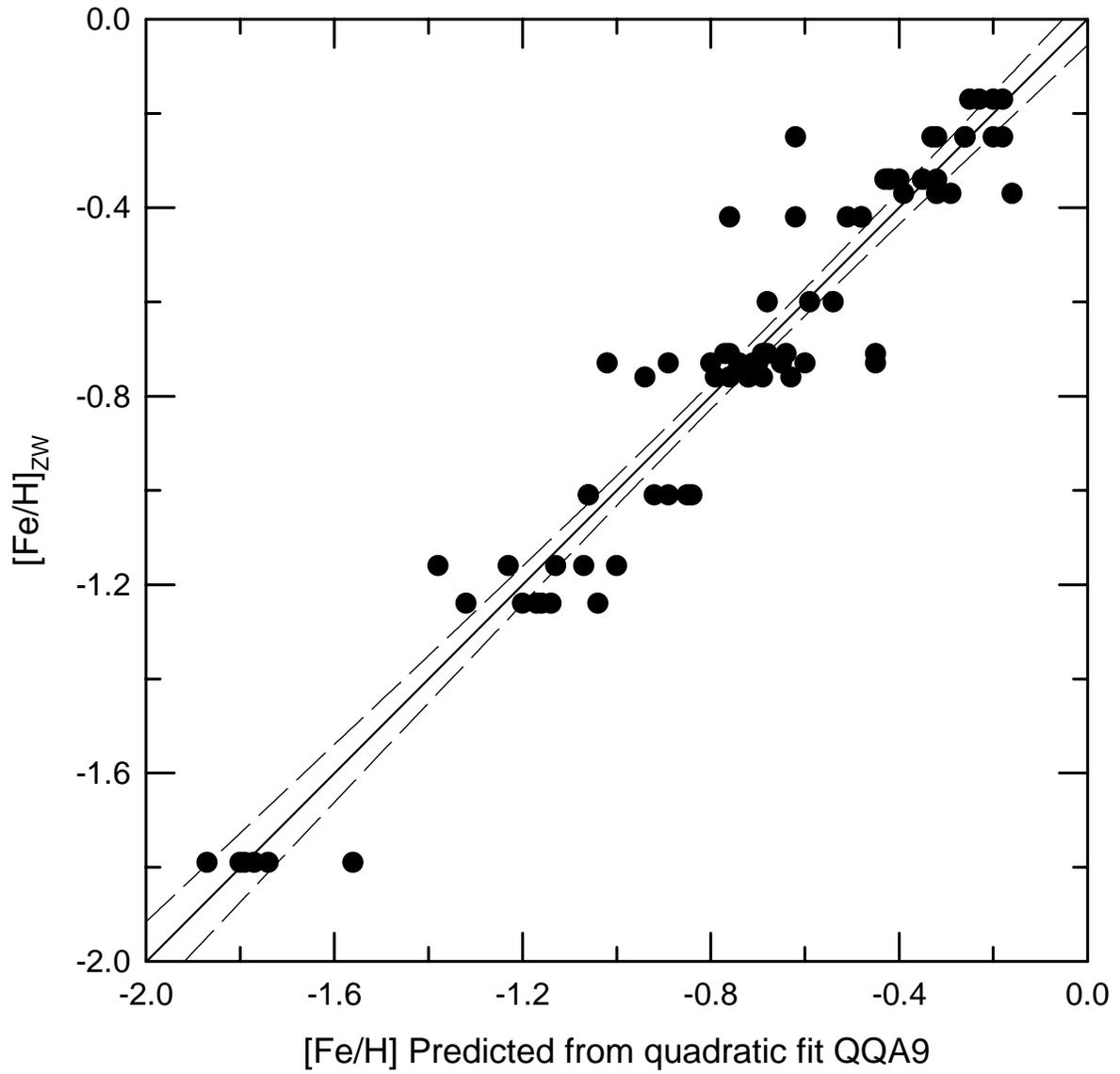

Fig. 10. — This is the same as Fig. 9 except that the fit was done to the reduced sample of 77 stars that also have colors and magnitudes (Table 10, solution QQA9).

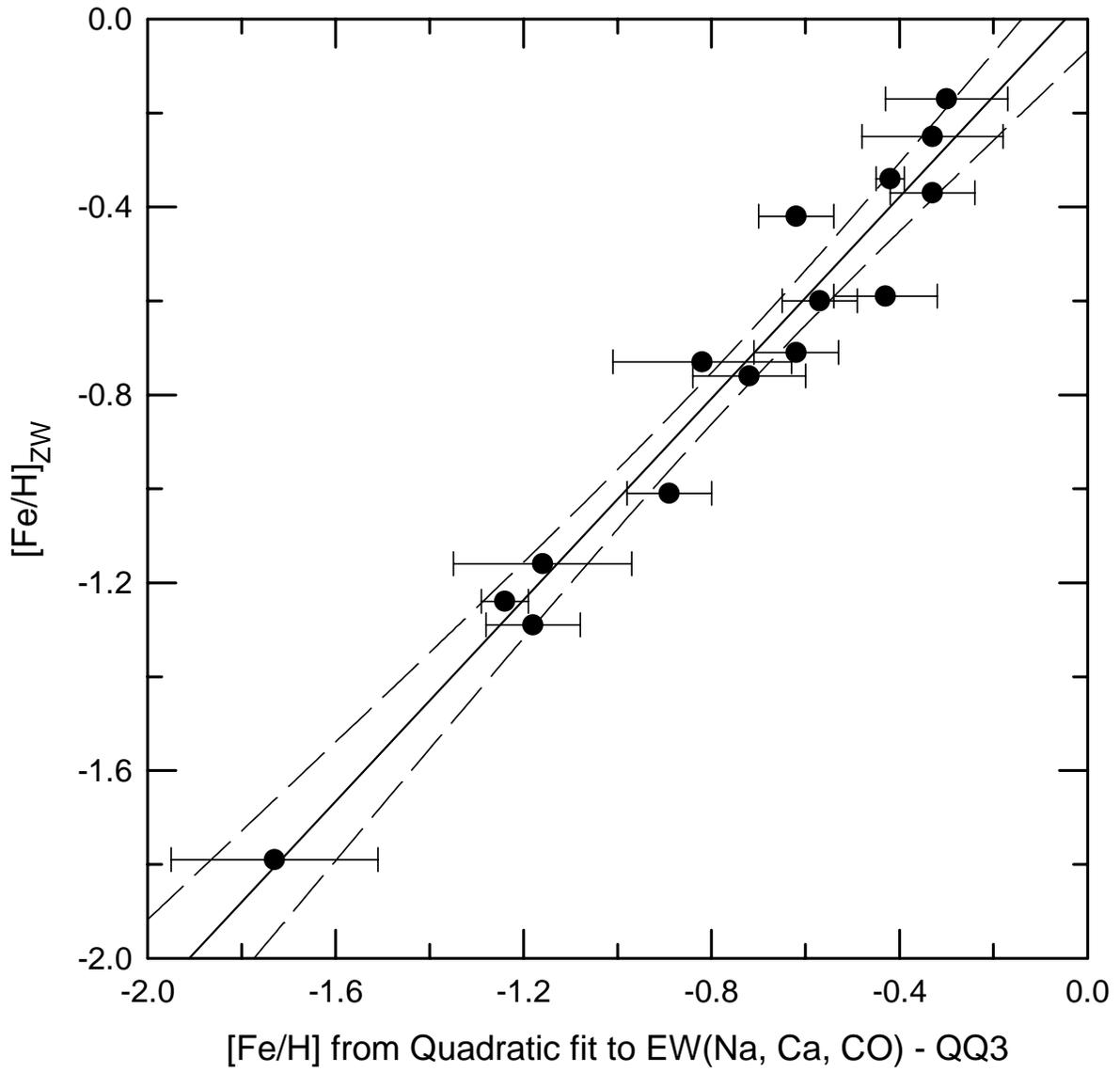

Fig. 11. — Same as Fig. 8 except that the fit to the 105 stars was a quadratic in the three spectroscopic indices (Table 10, solution QQ3) illustrated in Fig. 9).

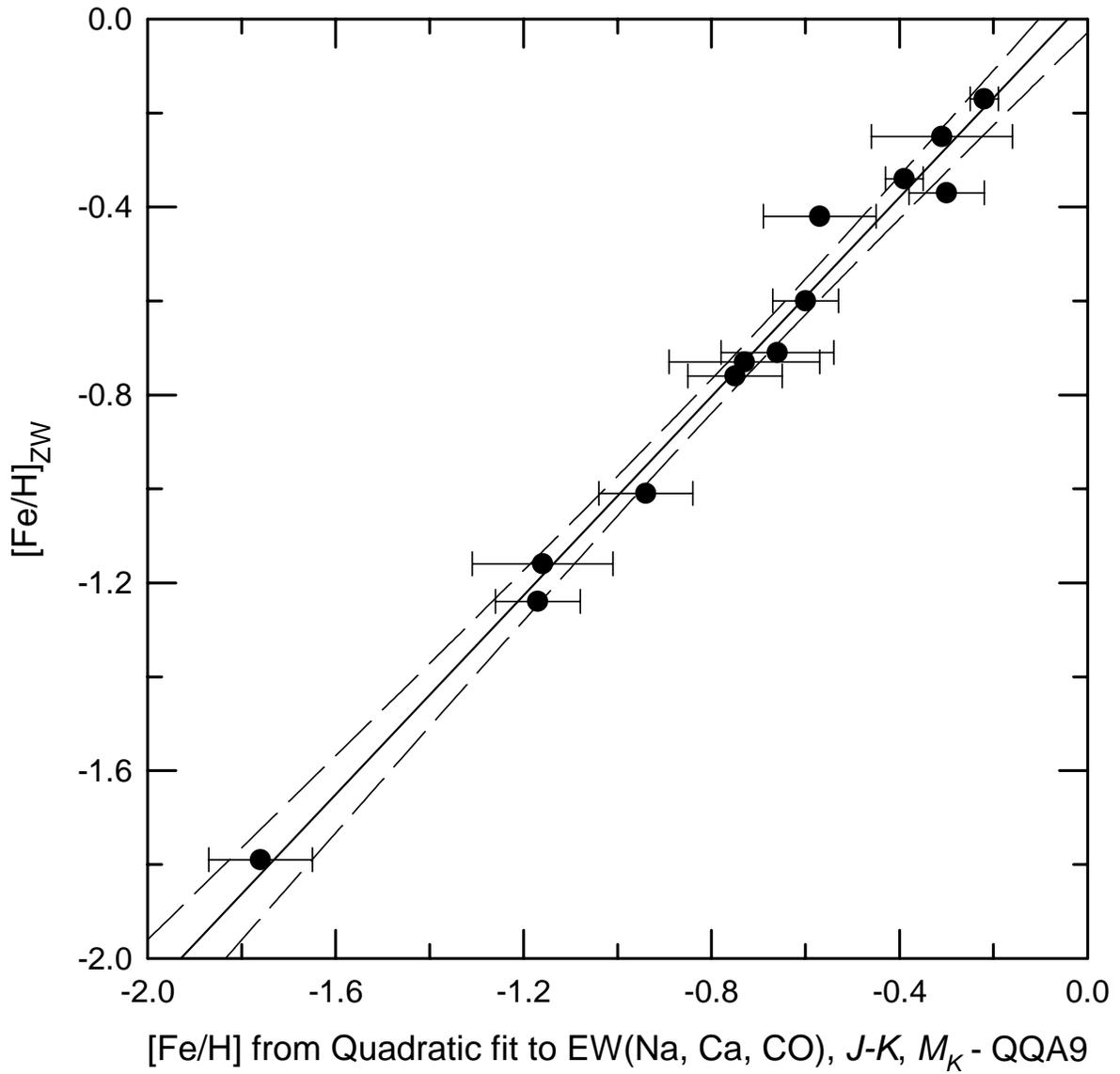

Fig. 12. — Same as Fig. 11 except for the quadratic fit found to both the spectroscopic and photometric indices for 77 stars (Table 10, solution QQA9, illustrated in Fig. 10).

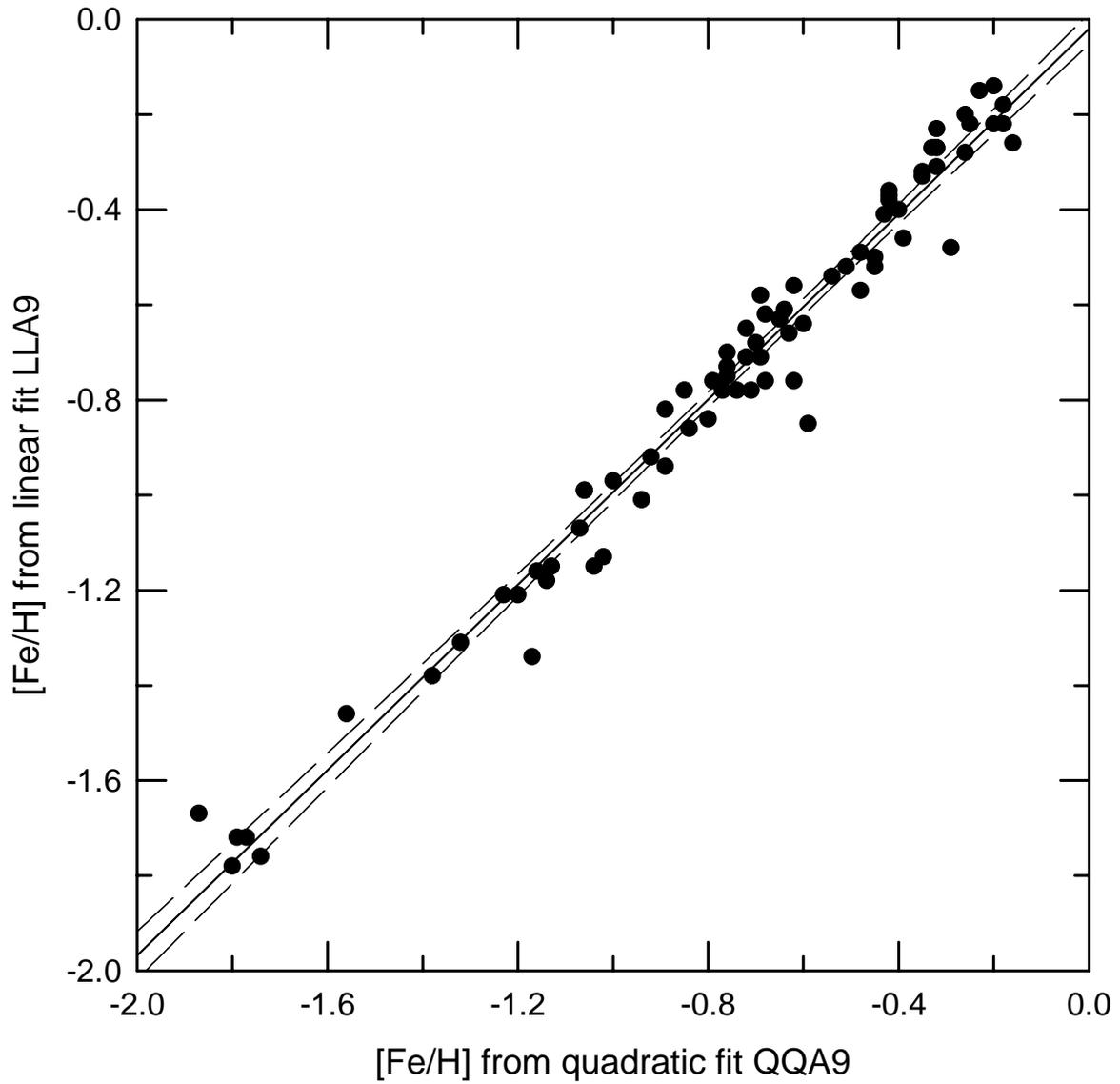

Fig. 13 - A comparison of linear and quadratic solutions, to both the three spectroscopic parameters and on $(J-K)_0$ and $M_K$. In both cases, the solutions given in Table 10 were applied to the same set of 77 stars and a value of [Fe/H] calculated from each one.

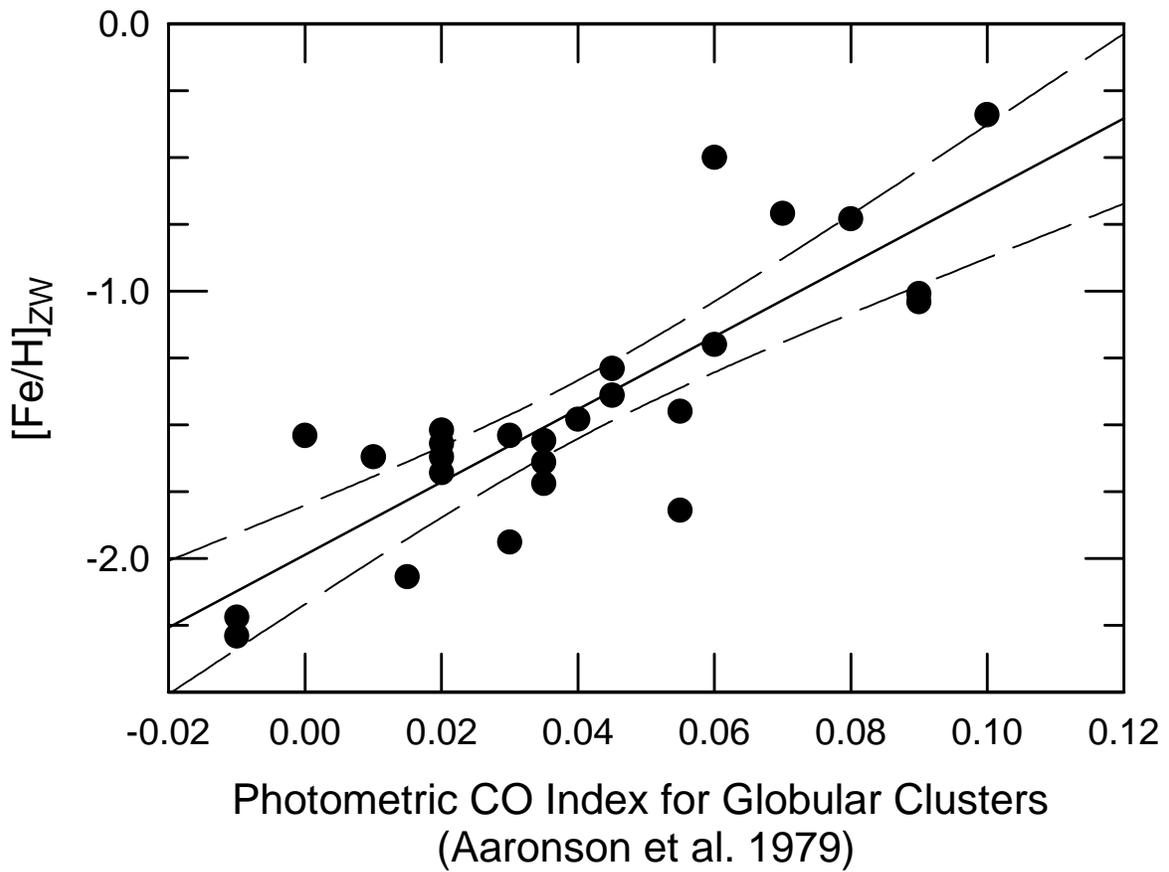

Fig. 14 - [Fe/H] on the ZW scale as updated by Harris (1996) is shown as a function of the photometric CO index measured in the integrated light of galactic globular clusters by Aaronson *et al.* (1978). The straight line is the least-squares fit to the data. The dashed lines are the 95% confidence limits.

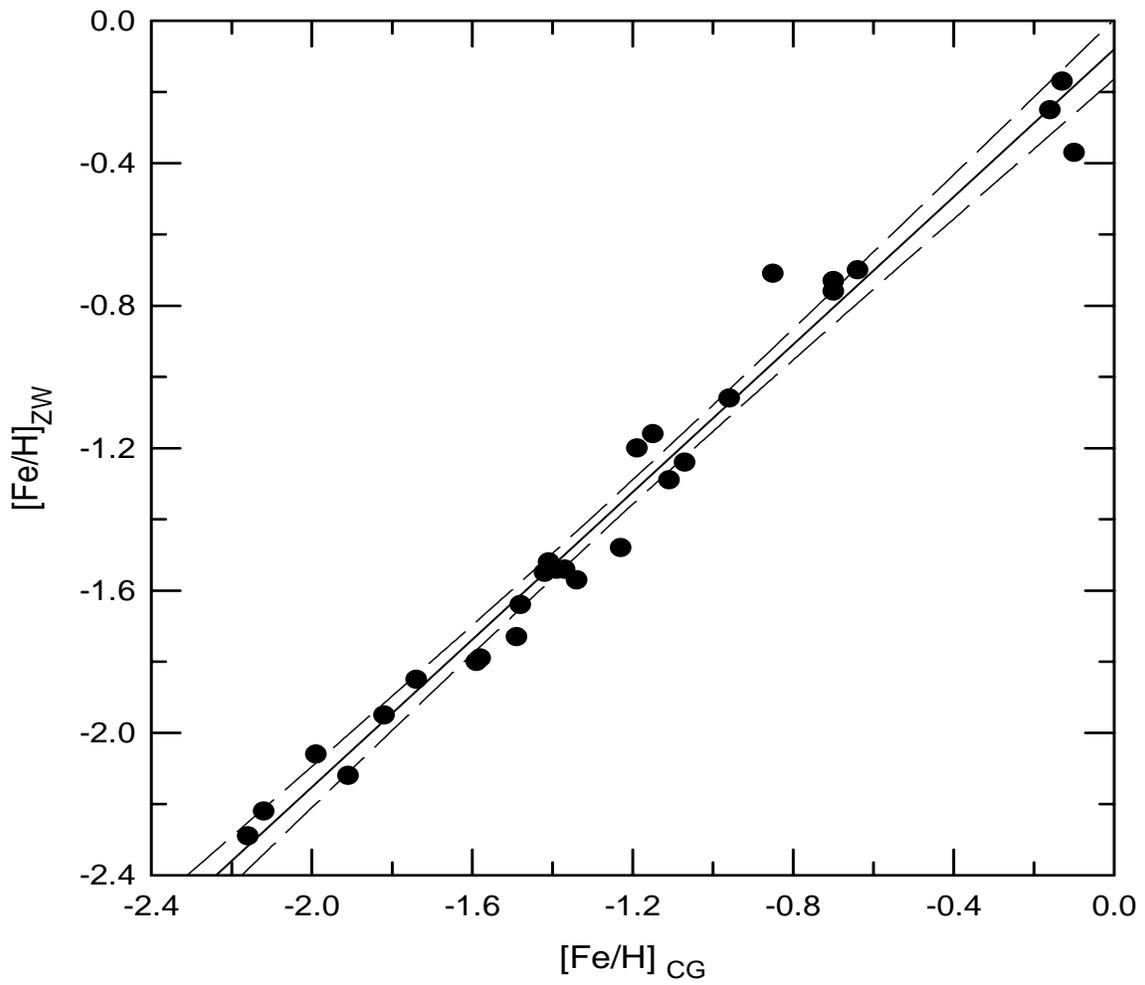

Fig. 15 - [Fe/H] on the ZW scale as updated by Harris (1996) is shown as a function of [Fe/H] from high resolution spectroscopy. The sources of the latter values are given in the text.

TABLE 1. Near-IR Spectral Indices, Colors, and Luminosities

| Cluster | Star | Photometric Parameters | | | | EW (Å) | | | Notes |
|---|---|---|---|---|---|---|---|---|---|
| | | $K_0$ | $(J-K)_0$ | $M_{K_0}$ | CO | Na | Ca | CO | |
| 47Tuc | A02=V21 | 6.74 | 1.21 | -6.43 | 0.160 | 2.09 | 1.63 | 18.08 | |
| 47Tuc | V08 | 6.66 | 1.12 | -6.51 | 0.130 | 1.42 | 2.32 | 9.22 | |
| 47Tuc | W12=V11 | 6.68 | 1.06 | -6.49 | 0.180 | 2.08 | 2.42 | 19.15 | |
| 47Tuc | A19 | 6.79 | 1.20 | -6.38 | 0.150 | 2.08 | 2.32 | 14.82 | |
| 47Tuc | V07 | 6.96 | 1.07 | -6.21 | 0.115 | 2.05 | 2.37 | 15.52 | |
| 47Tuc | V06 | 7.43 | 1.01 | -5.74 | 0.105 | 1.97 | 1.72 | 13.92 | |
| 47Tuc | L168=V18 | 7.44 | 1.00 | -5.73 | 0.160 | 1.77 | 1.55 | 18.14 | |
| 47Tuc | 5529 | 7.91 | 0.95 | -5.26 | 0.160 | 1.97 | 1.94 | 17.22 | faint |
| 47Tuc | 2426 | 8.48 | 0.90 | -4.69 | 0.110 | 1.74 | 2.22 | 13.66 | faint |
| 47Tuc | 1505 | 8.51 | 0.90 | -4.66 | 0.130 | 1.52 | 1.95 | 15.08 | faint |
| 47Tuc | 4418 | 8.54 | 0.89 | -4.63 | 0.145 | 1.46 | 1.69 | 15.48 | faint |
| 47Tuc | 1510 | 8.75 | 0.86 | -4.42 | 0.120 | 1.57 | 1.58 | 13.44 | faint |
| 47Tuc | 2416 | 9.36 | 0.83 | -3.81 | 0.155 | 1.49 | 1.29 | 14.42 | faint |
| 47Tuc | 6408 | 9.73 | 0.76 | -3.44 | 0.110 | 1.55 | 2.18 | 12.94 | faint |
| NGC288 | A260 | 8.47 | 0.97 | -6.52 | 0.120 | 0.67 | 2.36 | 6.95 | |
| NGC288 | A96 | 9.31 | 0.90 | -5.64 | 0.110 | 0.58 | 0.92 | 10.95 | |
| NGC288 | A78 | 9.64 | 0.87 | -5.29 | 0.090 | 0.61 | 0.82 | 11.34 | |
| NGC288 | C20 | 9.74 | 0.86 | -5.19 | 0.045 | 0.43 | -0.56 | 8.13 | |
| NGC288 | A77 | 10.01 | 0.83 | -4.90 | 0.065 | 0.62 | 0.30 | 9.96 | |
| NGC288 | A245 | 10.52 | 0.74 | -4.33 | 0.070 | 0.87 | -0.18 | 8.78 | |
| NGC362 | III-11 | 8.71 | 0.93 | -5.89 | 0.120 | 1.06 | 1.16 | 12.80 | |
| NGC362 | IV-100 | 8.99 | 0.89 | -5.61 | 0.080 | 0.66 | 0.16 | 11.01 | |
| NGC362 | III-63 | 9.01 | 0.92 | -5.59 | 0.090 | 1.30 | 1.04 | 13.57 | |
| NGC362 | III-44 | 9.33 | 0.89 | -5.27 | 0.060 | 0.98 | 0.93 | 8.25 | |
| NGC362 | III-70 | 9.52 | 0.79 | -5.08 | 0.010 | 0.58 | 0.49 | 4.62 | |
| NGC362 | 1 | … | … | … | … | 0.63 | 1.37 | 5.34 | 2 |
| NGC362 | 2 | … | … | … | … | 1.57 | 1.24 | 13.15 | 2 |
| NGC362 | 3 | … | … | … | … | 1.44 | 1.57 | 8.38 | 2 |
| NGC362 | 4 | … | … | … | … | 1.42 | 0.79 | 8.92 | 2 |
| NGC4833 | MA75 | 8.10 | 0.76 | -5.75 | 0.010 | -0.41 | 0.05 | 2.04 | |
| NGC4833 | V9 | 8.10 | 0.64 | -5.75 | 0.000 | 0.37 | 0.91 | 2.96 | |
| NGC4833 | B55 | 8.14 | 0.76 | -5.71 | 0.010 | -0.09 | -0.04 | 1.80 | |
| NGC4833 | D75 | 8.24 | 0.77 | -5.61 | -0.020 | -0.35 | 1.49 | 0.07 | |
| NGC4833 | MA1 | 8.26 | 0.73 | -5.59 | 0.020 | 0.74 | 1.41 | 4.13 | |
| NGC4833 | V16 | 8.42 | 0.71 | -5.43 | -0.030 | 0.10 | 1.79 | -1.60 | |

|  |  | Photometric Parameters | | | | EW (Å) | | | Notes |
|---|---|---|---|---|---|---|---|---|---|
| Cluster | Star | $K_0$ | $(J-K)_0$ | $M_{K_0}$ | CO | Na | Ca | CO | |
| NGC5927 | 100 | 7.77 | 1.02 | -6.58 | 0.240 | 3.88 | 3.66 | 21.07 | |
| NGC5927 | 799 | 8.52 | 1.02 | -5.83 | 0.210 | 3.38 | 2.95 | 20.75 | |
| NGC5927 | 627 | 8.78 | 1.01 | -5.57 | 0.175 | 3.40 | 4.08 | 16.98 | |
| NGC5927 | 532 | 9.34 | 0.94 | -5.01 | 0.185 | 3.04 | 3.41 | 13.34 | |
| NGC5927 | 622 | 9.92 | 0.85 | -4.43 | 0.120 | 2.92 | 4.17 | 11.79 | |
| NGC5927 | 536 | 10.55 | 0.79 | -3.80 | 0.195 | 1.61 | 1.95 | 15.78 | 1 |
| NGC6304 | 1 | … | … | … | … | 3.35 | 2.87 | 15.95 | |
| NGC6304 | 2 | … | … | … | … | 3.59 | 2.83 | 15.38 | |
| NGC6304 | 3 | … | … | … | … | 2.57 | 2.27 | 18.51 | |
| NGC6304 | 4 | … | … | … | … | 4.16 | 3.22 | 24.79 | |
| NGC6304 | 5 | 8.80 | … | -5.09 | … | 2.98 | 1.52 | 18.81 | |
| NGC6304 | 6 | … | … | … | … | 4.67 | 3.07 | 23.52 | |
| NGC6304 | 7 | 8.30 | … | -5.59 | … | 2.85 | 1.50 | 19.88 | |
| NGC6388 | 1=V2 | 8.10 | 0.96 | -7.20 | 0.200 | 2.07 | 0.92 | 20.36 | |
| NGC6388 | 2 | … | … | … | … | 2.32 | 2.04 | 19.17 | |
| NGC6388 | 3=V1 | 8.59 | 0.99 | -6.71 | 0.110 | 2.09 | 1.15 | 19.59 | |
| NGC6388 | 4=V3 | 8.90 | 1.11 | -6.40 | 0.140 | 2.68 | 2.54 | 17.12 | |
| NGC6388 | 5 | … | … | … | … | 2.64 | 2.44 | 19.52 | |
| NGC6388 | 6 | … | … | … | … | 2.63 | 2.41 | 22.19 | |
| NGC6388 | 7 | … | … | … | … | 1.63 | 1.23 | 26.65 | |
| NGC6388 | 8 | … | … | … | … | 2.69 | 2.77 | 14.23 | |
| NGC6388 | 9 | … | … | … | … | 2.90 | 2.08 | 16.35 | |
| NGC6440 | KF-1 | 8.25 | 1.02 | -5.75 | … | 3.42 | 3.48 | 14.62 | |
| NGC6440 | KF-2 | 8.30 | 1.04 | -5.70 | … | 3.04 | 3.13 | 17.70 | |
| NGC6440 | KF-3 | 8.30 | 1.10 | -5.70 | … | 3.02 | 3.45 | 16.26 | |
| NGC6440 | KF-4 | 8.36 | 1.12 | -5.64 | … | 3.47 | 3.02 | 16.01 | |
| NGC6440 | KF-5 | 8.47 | 1.01 | -5.53 | … | 2.84 | 2.77 | 20.46 | |
| NGC6440 | KF-6 | 8.52 | 1.08 | -5.48 | … | 3.28 | 3.35 | 15.47 | |
| NGC6440 | KF-8 | 8.77 | 1.08 | -5.23 | … | 2.99 | 3.09 | 16.33 | |
| NGC6440 | KF-7 | 8.77 | 1.05 | -5.23 | … | 2.66 | 2.94 | 16.80 | |
| NGC6528 | 24 | … | … | … | … | 2.80 | 2.70 | 16.90 | |
| NGC6528 | 2 | … | … | … | … | 5.07 | 3.13 | 24.23 | |
| NGC6528 | 8 | … | … | … | … | 2.70 | 1.48 | 23.94 | |
| NGC6528 | 1 | … | … | … | … | 3.95 | 4.24 | 20.88 | |
| NGC6528 | 7 | 8.36 | 1.01 | -6.01 | … | 3.95 | 4.10 | 18.66 | |
| NGC6528 | 11 | 7.87 | 1.09 | -6.49 | … | 3.72 | 4.20 | 20.68 | |
| NGC6528 | 22 | 8.52 | 1.04 | -5.84 | … | 3.51 | 3.20 | 23.67 | |
| NGC6528 | 6 | 8.57 | 0.98 | -5.79 | … | 3.65 | 3.32 | 21.26 | |
| NGC6528 | 5 | … | … | … | … | 3.58 | 3.30 | 17.62 | |

| Cluster | Star | Photometric Parameters | | | | EW (Å) | | | Notes |
|---|---|---|---|---|---|---|---|---|---|
| | | $K_0$ | $(J-K)_0$ | $M_{K_0}$ | CO | Na | Ca | CO | |
| NGC6553 | 20=v4 | 6.05 | 1.06 | -6.91 | 0.200 | 2.30 | 0.77 | 17.57 | |
| NGC6553 | 19 | 6.74 | 1.20 | -6.62 | … | 3.88 | 3.34 | 19.31 | |
| NGC6553 | 25 | 7.06 | 1.08 | -6.30 | … | 3.59 | 3.55 | 19.44 | |
| NGC6553 | 16 | 7.28 | 0.98 | -6.08 | … | 3.45 | 3.89 | 24.84 | |
| NGC6553 | 26=v5 | 6.18 | 1.02 | -5.96 | 0.210 | 3.28 | 2.62 | 24.20 | |
| NGC6553 | 14=V7 | 7.53 | 1.04 | -5.83 | … | 3.60 | 3.87 | 22.77 | |
| NGC6553 | 2=IV24=V13 | 7.59 | 1.14 | -5.77 | … | 3.59 | 2.66 | 15.88 | |
| NGC6553 | 18 | … | … | … | … | 3.82 | 3.40 | 15.87 | |
| NGC6553 | 5 | … | … | … | … | 3.62 | 4.05 | 20.66 | |
| NGC6624 | KF-1 | 8.59 | 0.98 | -5.81 | … | 2.33 | 2.24 | 19.53 | |
| NGC6624 | KF-2 | 8.90 | 0.98 | -5.50 | … | 3.14 | 2.73 | 12.65 | |
| NGC6624 | KF-3 | 8.92 | 0.98 | -5.48 | … | 2.07 | 1.95 | 12.81 | |
| NGC6624 | KF-4 | 9.43 | 0.99 | -4.97 | … | 2.22 | 1.82 | 18.42 | |
| NGC6624 | KF-5 | 9.75 | 0.90 | -4.65 | … | 2.14 | 2.04 | 16.44 | |
| NGC6712 | LM5 | 8.71 | 0.88 | -5.52 | … | 1.41 | 1.42 | 18.89 | |
| NGC6712 | LCO1 | 8.80 | 0.87 | -5.43 | … | 1.70 | 1.43 | 17.79 | |
| NGC6712 | LCO3 | 8.88 | 0.83 | -5.35 | … | 1.11 | 0.95 | 15.51 | |
| NGC6712 | LM8 | 9.07 | 0.88 | -5.16 | … | 1.49 | 1.35 | 12.28 | |
| NGC6712 | LM10 | 9.09 | 0.81 | -5.14 | … | 1.01 | 0.46 | 16.12 | |
| NGC6712 | B66 | 9.39 | 0.84 | -4.84 | … | 1.25 | 1.01 | 16.16 | |
| M5 | 1 | … | … | … | … | 1.04 | 1.69 | 12.23 | 3 |
| M5 | 2 | … | … | … | … | 0.12 | 2.14 | 12.39 | 3 |
| M5 | 3 | … | … | … | … | 0.77 | 1.88 | 8.66 | 3 |
| M5 | 4 | … | … | … | … | 0.89 | 0.27 | 7.23 | 3 |
| M5 | 5 | … | … | … | … | 0.29 | 3.19 | 10.15 | 3 |
| M5 | 6 | … | … | … | … | 1.29 | 0.15 | 10.21 | 3 |
| M5 | 7 | … | … | … | … | 0.45 | 1.60 | 12.37 | 3 |
| M5 | V50 | … | … | … | … | 0.57 | 1.51 | 10.35 | 3 |
| M15 | 1 | … | … | … | … | -0.75 | -0.03 | 2.42 | 3 |
| M15 | 2 | … | … | … | … | 0.73 | 0.87 | 3.19 | 3 |
| M15 | 3 | … | … | … | … | 0.35 | -0.87 | 2.09 | 3 |
| M15 | 4 | 7.91 | 0.95 | -5.26 | 0.160 | -0.17 | -0.70 | 1.63 | 3 |
| M15 | 5 | 8.48 | 0.90 | -4.69 | 0.110 | -0.40 | 0.26 | 0.77 | 3 |
| M15 | 6 | 8.51 | 0.90 | -4.66 | 0.130 | -0.80 | 0.57 | 2.05 | 3 |
| M15 | 7 | 8.54 | 0.89 | -4.63 | 0.145 | 0.05 | 0.19 | 1.65 | 3 |

|  |  | Photometric Parameters | | | | EW (Å) | | | Notes |
| --- | --- | --- | --- | --- | --- | --- | --- | --- | --- |
| Cluster | Star | $K_0$ | $(J-K)_0$ | $M_{K_0}$ | CO | Na | Ca | CO | |
| M69 | 1=In-IV-11=V8 | 8.04 | 1.03 | -6.53 | … | 3.06 | 2.49 | 18.12 | |
| M69 | II-37=V6 | 8.18 | 1.10 | -6.39 | 0.160 | 2.22 | 2.41 | 14.48 | |
| M69 | I-40=V3 | 8.46 | 0.98 | -6.11 | 0.140 | 2.38 | 1.59 | 13.24 | |
| M69 | 2=In-I-12=V1 | 8.54 | 1.01 | -6.03 | 0.195 | 1.68 | 2.42 | 18.15 | |
| M69 | 3=In-IV-27 | 8.54 | 1.01 | -6.03 | … | 2.39 | 2.19 | 17.83 | |
| M69 | 4 | 8.78 | 0.99 | -5.80 | … | 2.14 | 2.09 | 18.49 | |
| M71 | 29 | … | 1.15 | -6.42 | 0.15 | 2.06 | 2.87 | 16.96 | |
| M71 | 30 | … | 0.94 | -5.17 | 0.16 | 0.61 | 2.87 | 15.17 | |
| M71 | B | … | 1.07 | -6.12 | 0.11 | 2.10 | 2.55 | 15.51 | |
| M71 | 46 | … | 0.93 | -4.97 | 0.15 | 1.27 | 2.00 | 16.23 | |
| M71 | A4 | … | 0.86 | -4.94 | 0.13 | 1.36 | 2.63 | 14.52 | |
| M71 | 1=H | 8.05 | 1.05 | -4.85 | … | 2.42 | 1.64 | 14.17 | |
| M71 | 2=I | 8.64 | 0.69 | -4.26 | … | 0.59 | 1.10 | 11.45 | |
| M71 | 3=113 | 7.96 | 0.96 | -4.94 | 0.165 | 1.88 | 1.84 | 15.88 | |
| M71 | 4=45 | 8.05 | 0.93 | -4.85 | 0.110 | 1.56 | 1.52 | 13.82 | |
| M71 | 5=64 | 9.31 | 0.86 | -3.59 | … | 0.81 | 1.80 | 12.53 | 1 |
| M71 | 6=66 | 9.34 | 0.66 | -3.56 | … | 1.23 | 1.10 | 10.22 | 1 |
| M71 | 7=36 | 9.31 | … | -3.59 | … | 1.27 | 1.35 | 10.61 | 1 |
| M71 | 8=21 | 9.47 | 0.73 | -3.43 | 0.105 | 0.89 | 0.32 | 10.34 | 1 |
| M92 | 1 | … | … | … | … | -0.30 | 0.62 | 0.88 | 3 |
| M92 | 2 | … | … | … | … | -0.11 | 0.07 | 0.93 | 3 |
| M92 | 3 | … | … | … | … | 0.20 | 0.32 | 0.25 | 3 |
| M92 | 4 | … | … | … | … | 0.37 | 0.21 | 0.99 | 3 |
| M92 | 5 | … | … | … | … | 0.02 | 0.09 | 1.44 | 3 |

Notes to Table 2

1. $M_K$ too faint for inclusion in calibration sample
2. These are 4 bright stars in the center of NGC 362 used as a test. They are not part of the calibration sample
3. These are bright stars in the central part of the cluster with no previous observations.

TABLE 2. Equivalent Width Measurement Intervals

| Feature | Wavelengths (μm) |
|---|---|
| Na I | 2.2040 - 2.2107 |
| Na I continuum | 2.1910 - 2.1966 |
| Na I continuum | 2.2125 - 2.2170 |
| Ca I | 2.2577 - 2.2692 |
| Ca I continuum | 2.2450 - 2.2560 |
| Ca I continuum | 2.2700 - 2.2720 |
| $^{12}CO(2,0)$ band | 2.2910 - 2.3020 |
| $^{12}CO(2,0)$ continuum | 2.2300 - 2.2370 |
| $^{12}CO(2,0)$ continuum | 2.2420 - 2.2580 |
| $^{12}CO(2,0)$ continuum | 2.2680 - 2.2790 |
| $^{12}CO(2,0)$ continuum | 2.2840 - 2.2910 |

TABLE 3. Mean Differences in EW Measurements of 17 Stars

|  | EW(Na) | EW(Ca) | EW(CO) |
|---|---|---|---|
| Mean difference | 0.34 | 0.90 | 0.90 |
| Dispersion | 0.28 | 0.62 | 0.74 |

TABLE 4. Cluster Mean Values and Dispersions for Observed Parameters

| Cluster | [Fe/H] (Harris) | EW(Na) mean | σ | EW(Ca) mean | σ | EW(CO) mean | σ | $(J-K)_0$ mean | σ | $M_{K0}$ mean | σ |
|---|---|---|---|---|---|---|---|---|---|---|---|
| 47 Tuc | -0.76 | 1.92 | 0.25 | 2.05 | 0.39 | 15.55 | 3.40 | 1.10 | 0.08 | -6.21 | 0.34 |
| NGC 288 | -1.24 | 0.63 | 0.14 | 0.61 | 1.03 | 9.35 | 1.70 | 0.86 | 0.08 | -5.31 | 0.73 |
| NGC 362 | -1.16 | 0.91 | 0.30 | 0.75 | 0.42 | 10.05 | 3.66 | 0.88 | 0.06 | -5.48 | 0.32 |
| NGC4833 | -1.79 | 0.06 | 0.44 | 0.93 | 0.77 | 1.56 | 2.05 | 0.73 | 0.05 | -5.64 | 0.12 |
| NGC5927 | -0.37 | 3.32 | 0.37 | 3.65 | 0.50 | 16.79 | 4.21 | 0.97 | 0.07 | -5.48 | 0.82 |
| NGC6304 | -0.59 | 3.45 | 0.75 | 2.47 | 0.72 | 19.55 | 3.54 | … | … | -5.34 | 0.35 |
| NGC6388 | -0.60 | 2.40 | 0.41 | 1.95 | 0.68 | 19.46 | 3.58 | 1.02 | 0.08 | -6.77 | 0.40 |
| NGC6440 | -0.34 | 3.09 | 0.28 | 3.15 | 0.25 | 16.71 | 1.77 | 1.06 | 0.04 | -5.53 | 0.21 |
| NGC6528 | -0.17 | 3.66 | 0.69 | 3.30 | 0.87 | 20.87 | 2.74 | 1.03 | 0.05 | -6.03 | 0.32 |
| NGC6553 | -0.25 | 3.46 | 0.47 | 3.13 | 1.02 | 20.06 | 3.35 | 1.07 | 0.07 | -6.21 | 0.42 |
| NGC6624 | -0.42 | 2.38 | 0.43 | 2.15 | 0.36 | 15.97 | 3.16 | 0.96 | 0.04 | -5.28 | 0.46 |
| NGC6712 | -1.01 | 1.33 | 0.26 | 1.10 | 0.38 | 16.12 | 2.26 | 0.85 | 0.03 | -5.24 | 0.25 |
| M5 | -1.29 | 0.68 | 0.39 | 1.55 | 0.98 | 10.45 | 1.87 | … | … | … | … |
| M15 | -2.22 | -0.14 | 0.56 | 0.04 | 0.64 | 1.97 | 0.75 | … | … | … | … |
| M69 | -0.71 | 2.31 | 0.45 | 2.20 | 0.34 | 16.72 | 2.26 | 1.02 | 0.04 | -6.15 | 0.27 |
| M71 | -0.73 | 1.54 | 0.65 | 2.11 | 0.64 | 14.86 | 1.63 | 0.95 | 0.13 | -5.17 | 0.67 |
| M92 | -2.29 | 0.04 | 0.26 | 0.26 | 0.22 | 0.90 | 0.43 | … | … | … | … |

TABLE 5. PHOTOMETRIC AND SPECTROSCOPIC VARIABLES

| Variable | Number of Stars | Mean | Standard Deviation |
|---|---|---|---|
| $(J-K)_0$ | 77 | 0.965 | 0.124 |
| $M_K$ | 77 | -5.68 | 0.61 |
| [Fe/H] | 105 | -0.728 | 0.436 |
| EW(Na) | 105 | 2.155 | 1.221 |
| EW(Ca) | 105 | 2.148 | 1.108 |
| EW(CO) | 105 | 15.40 | 5.55 |

TABLE 6. EIGENVALUES FOR PCA WITH
SPECTROSCOPIC AND PHOTOMETRIC VARIABLES

| Eigenvectors | Eigenvalue | Percent Variance | Cumulative Percent |
|---|---|---|---|
| V1 | 4.113 | 68.5 | 68.5 |
| V2 | 0.988 | 16.5 | 85.0 |
| V3 | 0.457 | 7.6 | 92.6 |
| V4 | 0.257 | 4.3 | 96.9 |
| V5 | 0.122 | 2.0 | 99.0 |

TABLE 7. Eigenvectors for PCA with Spectroscopic
and Photometric Variables

| Normalized Variable | V1 | V2 | V3 | V4 | V5 |
|---|---|---|---|---|---|
| EW(Na) | -.947 | -.157 | .083 | -.067 | .213 |
| EW(Ca) | -.827 | -.210 | .482 | -.108 | -.167 |
| EW(CO) | -.853 | -.097 | -.450 | -.179 | -.162 |
| $(J-K)_0$ | -.855 | .303 | -.019 | .413 | -.072 |
| $M_K$ | .461 | -.863 | -.053 | .190 | -.044 |
| [Fe/H] | -.928 | -.271 | -.110 | .043 | .124 |

TABLE 8. Eigenvalues for PCA with
Spectroscopic Variables Only

| Eigenvectors | Eigenvalue | Percent Variance | Cumulative Percent |
|---|---|---|---|
| V1 | 3.233 | 80.8 | 80.8 |
| V2 | .489 | 12.2 | 93.1 |
| V3 | .168 | 4.2 | 97.3 |

TABLE 9. Eigenvectors for PCA with
Spectroscopic Variables

| Variable | V1 | V2 | V3 |
|---|---|---|---|
| EW(Na) | 0.950 | 0.040 | -0.220 |
| EW(Ca) | 0.828 | 0.526 | 0.192 |
| EW(CO) | 0.859 | -0.451 | 0.237 |
| [Fe/H] | 0.952 | -0.091 | -0.162 |

TABLE 10. The Linear and Quadratic Solutions for [Fe/H]

| variable | Linear (LL1) | | Linear (LL3) | | Linear (LLA9) | | Quadratic (QQ1) | | Quadratic (QQ3) | | Quadratic (QQA9) | |
|---|---|---|---|---|---|---|---|---|---|---|---|---|
| | | | coeff. | $\sigma$ | coeff. | $\sigma$ | | | coeff. | $\sigma$ | coeff. | $\sigma$ |
| const | -1.706 | 0.075 | -1.663 | 0.057 | -1.451 | 0.18 | -1.935 | 0.11 | -1.811 | 0.074 | 1.097 | 1.16 |
| EW(Na) | … | … | 0.182 | 0.029 | 0.202 | 0.036 | … | … | 0.389 | 0.065 | 0.130 | 0.091 |
| EW(Na)$^2$ | … | … | … | … | … | … | … | … | -0.047 | 0.013 | 0.016 | 0.020 |
| EW(Ca) | … | … | 0.057 | 0.025 | 0.025 | 0.026 | … | … | -0.030 | 0.051 | 0.058 | 0.056 |
| EW(Ca)$^2$ | … | … | … | … | … | … | … | … | 0.024 | 0.012 | -0.004 | 0.014 |
| EW(CO) | 0.063 | 0.005 | 0.027 | 0.005 | 0.026 | 0.005 | 0.106 | 0.015 | 0.043 | 0.013 | 0.028 | 0.016 |
| EW(CO)$^2$ | … | … | … | … | … | … | -0.002 | 0.001 | -0.001 | 0.000 | 0.000 | 0.001 |
| $(J-K)_0$ | … | … | … | … | 0.749 | 0.23 | … | … | … | … | 5.240 | 1.84 |
| $(J-K)_0^2$ | … | … | … | … | … | … | … | … | … | … | -2.313 | 0.93 |
| $M_K$ | … | … | … | … | 0.151 | 0.033 | … | … | … | … | 1.812 | 0.40 |
| $M_K^2$ | … | … | … | … | … | … | … | … | … | … | 0.147 | 0.034 |
| R | 0.81 | | 0.91 | | 0.95 | | 0.82 | | 0.94 | | 0.96 | |
| $\sigma$ (est) | 0.17 | | 0.11 | | 0.07 | | 0.18 | | 0.09 | | 0.06 | |

TABLE 11. Input and Output [Fe/H] for the Calibration Clusters

| Cluster | [Fe/H] (Harris) | Stars used in solutions | | Linear (LL3) | | Linear (LLA9) | | Quadratic (QQ1) | | Quadratic (QQ3) | | Quadratic (QQA9) | |
|---|---|---|---|---|---|---|---|---|---|---|---|---|---|
| | | all | +phot. | [Fe/H] | disp. | [Fe/H] | disp. | [Fe/H] | disp. | [Fe/H] | disp. | [Fe/H] | disp. |
| 47 Tuc | -0.76 | 7 | 7 | -0.77 | ±0.13 | -0.73 | ±0.14 | -0.69 | ±0.21 | -0.72 | ±0.12 | -0.75 | ±0.10 |
| NGC 288 | -1.24 | 6 | 6 | -1.26 | 0.07 | -1.22 | 0.08 | -1.09 | 0.13 | -1.24 | 0.05 | -1.17 | 0.09 |
| NGC 362 | -1.16 | 5 | 5 | -1.18 | 0.16 | -1.15 | 0.16 | -1.05 | 0.28 | -1.16 | 0.19 | -1.16 | 0.15 |
| NGC4833 | -1.79 | 6 | 6 | -1.56 | 0.13 | -1.68 | 0.12 | -1.78 | 0.21 | -1.73 | 0.22 | -1.76 | 0.11 |
| NGC5927 | -0.37 | 5 | 5 | -0.39 | 0.17 | -0.36 | 0.11 | -0.63 | 0.23 | -0.33 | 0.09 | -0.30 | 0.08 |
| NGC6304 | -0.59 | 7 | 0 | -0.36 | 0.24 | | | -0.49 | 0.15 | -0.43 | 0.11 | | |
| NGC6388 | -0.60 | 9 | 3 | -0.58 | 0.08 | -0.72 | 0.16 | -0.53 | 0.15 | -0.57 | 0.08 | -0.60 | 0.07 |
| NGC6440 | -0.34 | 8 | 8 | -0.47 | 0.04 | -0.36 | 0.05 | -0.61 | 0.09 | -0.42 | 0.03 | -0.39 | 0.04 |
| NGC6528 | -0.17 | 9 | 4 | -0.24 | 0.18 | -0.20 | 0.03 | -0.43 | 0.11 | -0.30 | 0.13 | -0.22 | 0.03 |
| NGC6553 | -0.25 | 9 | 7 | -0.31 | 0.18 | -0.29 | 0.21 | -0.46 | 0.14 | -0.33 | 0.15 | -0.31 | 0.15 |
| NGC6624 | -0.42 | 5 | 5 | -0.67 | 0.10 | -0.58 | 0.10 | -0.66 | 0.18 | -0.62 | 0.08 | -0.57 | 0.12 |
| NGC6712 | -1.01 | 6 | 6 | -0.92 | 0.10 | -0.89 | 0.09 | -0.65 | 0.13 | -0.89 | 0.09 | -0.94 | 0.10 |
| M5 | -1.29 | 9 | 0 | -1.17 | 0.07 | | | -1.01 | 0.14 | -1.18 | 0.10 | | |
| M69 | -0.71 | 6 | 6 | -0.66 | 0.11 | -0.66 | 0.10 | -0.61 | 0.13 | -0.62 | 0.09 | -0.66 | 0.12 |
| M71 | -0.73 | 4 | 4 | -0.86 | 0.16 | -0.77 | 0.19 | -0.72 | 0.10 | -0.82 | 0.19 | -0.73 | 0.16 |
| ⟨fit − ZW⟩ | | | | 0.00 | | −0.01 | | 0.00 | | 0.00 | | 0.00 | |
| σ⟨fit − ZW⟩ | | | | 0.13 | | 0.08 | | 0.20 | | 0.10 | | 0.06 | |
| Average | | | | | 0.13 | | 0.12 | | 0.16 | | 0.11 | | 0.10 |
| R | | | | 0.91 | | 0.95 | | 0.82 | | 0.94 | | 0.96 | |

14